\documentclass[10pt,draftcls,onecolumn]{IEEEtran}
\usepackage{ifpdf, flushend}
\ifCLASSINFOpdf
  \usepackage[pdftex]{graphicx}
  \graphicspath{{../pdf/}{../jpeg/}}
  \DeclareGraphicsExtensions{.pdf,.jpeg,.png}
\else
  \usepackage[dvips]{graphicx}
  \graphicspath{{../eps/}}
  \DeclareGraphicsExtensions{.eps}
\fi
\usepackage[cmex10]{amsmath}
\usepackage {amssymb}
\usepackage{algorithmic}
\usepackage{array}
\usepackage{mdwmath}
\usepackage{mdwtab}
\usepackage{eqparbox}
\usepackage{hyperref}
\usepackage{algorithm}
\usepackage{algorithmic}

\newcommand{\argmax}{\operatornamewithlimits{argmax}}
\newcommand{\beq}{\begin{equation}}
\newcommand{\eeq}{\end{equation}}
\newcommand{\beqn}{\begin{eqnarray}}
\newcommand{\eeqn}{\end{eqnarray}}
\newcommand{\beqno}{\begin{eqnarray*}}
\newcommand{\eeqno}{\end{eqnarray*}}
\newcommand{\bma}{\begin{displaymath}}
\newcommand{\ema}{\end{displaymath}}
\newcommand{\bnu}{\begin{enumerate}}
\newcommand{\enu}{\end{enumerate}}
\newcommand{\bce}{\begin{center}}
\newcommand{\ece}{\end{center}}
\newcommand{\btb}{\begin{tabular}}
\newcommand{\etb}{\end{tabular}}

\begin{document}
%
% paper title
% can use linebreaks \\ within to get better formatting as desired
\title{Joint Cooperative Spectrum Sensing and MAC Protocol Design for Multi-channel Cognitive Radio Networks}

%\author{Le~Thanh~Tan,~\IEEEmembership{Student Member,~IEEE,} Long~Bao~Le,~\IEEEmembership{Member,~IEEE} 
%\thanks{ Manuscript received April 08, 2011; revised June 24, 2011; accepted August 10, 2011. 
%The editor coordinating the review of this paper and approving it for publication is  Dr. Abbas Jamalipour. 
%}
%\thanks{L. T. Tan is with INRS-EMT, University of Quebec, Montr\'{e}al, Qu\'{e}bec, Canada. E-mail: lethanh@emt.inrs.ca. }
%\thanks{L. B. Le is with INRS-EMT, University of Quebec, Montr\'{e}al, Qu\'{e}bec, Canada. E-mail: long.le@emt.inrs.ca. } 
%}

\author{\IEEEauthorblockN{Le Thanh Tan and Long Bao Le}  %\vspace{-0.2cm}
%\author{Le~Thanh~Tan,~\IEEEmembership{Student Member,~IEEE} and Long~Bao~Le,~\IEEEmembership{Senior Member,~IEEE} 
\thanks{ Manuscript received January 16, 2014; revised April 19, 2014; accepted June 5, 2014. 
The editor coordinating the review of this paper and approving it for publication is  Dr. Ashish Pandharipande. 
}
\thanks{The authors are with the Institut National de la Recherche Scientifique--
\'{E}nergie, Mat\'{e}riaux et T\'{e}l\'{e}communications, Universit\'{e} du Qu\'{e}bec, Montr\'{e}al,
Qu\'{e}bec, QC J3X 1S2, Canada. 
Emails: lethanh@emt.inrs.ca; long.le@emt.inrs.ca. \textbf{L. T. Tan} is the corresponding author.}}

\markboth{EURASIP Journal on Wireless Communications and Networking} {Le \MakeLowercase{\textit{et al.}}: Joint Cooperative Spectrum Sensing and MAC Protocol Design for Multi-channel Cognitive Radio Networks}
\markright{EURASIP Journal on Wireless Communications and Networking} 

% make the title area
%\twocolumn[
\maketitle

%\begin{onecolabstract}
\begin{abstract}
\boldmath
In this paper, we propose a semi-distributed cooperative spectrum sensing (SDCSS) and channel access framework for multi-channel cognitive radio networks (CRNs). 
In particular, we consider a SDCSS scheme where secondary users (SUs) perform sensing and exchange sensing outcomes with each other to locate spectrum holes.
In addition, we devise the $p$-persistent CSMA-based cognitive MAC protocol integrating the SDCSS
to enable efficient spectrum sharing among SUs. We then perform throughput analysis and develop an algorithm to determine 
 the spectrum sensing and access parameters to maximize the throughput for a given allocation of channel sensing sets. Moreover, we consider the spectrum sensing 
set optimization problem for SUs to maximize the overall system throughput. We present both exhaustive search and low-complexity greedy algorithms
to determine the sensing sets for SUs and analyze their complexity. We also show how our design and analysis can be extended to consider
reporting errors. Finally, extensive numerical results are presented to demonstrate the significant performance gain of our optimized design framework
with respect to non-optimized designs as well as the impacts of different protocol parameters on the throughput performance.

\end{abstract}
%\end{onecolabstract}
%]

\IEEEpeerreviewmaketitle

\section{Introduction}

It has been well recognized that cognitive radio is one of the most important technologies that would enable us to meet 
exponentially growing spectrum demand via fundamentally improving the utilization of our precious spectral resources \cite{Zhao07}.
Development of efficient spectrum sensing and access algorithms for cognitive radios are among the key research issues for
successful deployment of this promising technology. 
There is indeed a growing literature on MAC protocol design and analysis for CRNs \cite{R1}-\cite{Do05} 
(see \cite{Cor09} for a survey of recent works in this topic). In \cite{R1}, it was shown that a significant throughput 
gain can be achieved by optimizing the sensing time under the single-SU setting.
Another related effort along this line was conducted in \cite{Kim08} where sensing-period optimization and optimal channel-sequencing 
algorithms were proposed to efficiently discover spectrum holes and to minimize the exploration delay.

In \cite{Su08}, a control-channel based MAC protocol was proposed for SUs to exploit white spaces in the cognitive ad hoc network. 
In particular, the authors of this paper developed both random and negotiation-based spectrum sensing schemes and performed throughput analysis 
for both saturation and non-saturation scenarios. There exists several other synchronous cognitive MAC protocols, which rely on a control channel 
for spectrum negotiation and access \cite{Su07}-\cite{Do05}, \cite{su081}.
In \cite{Le11} and \cite{Le12}, we designed, analyzed, and optimized a window-based MAC protocol to achieve efficient tradeoff between sensing time and contention overhead. 
However, these works considered the conventional single-user-energy-detection-based spectrum sensing scheme, which would only work well if the signal to noise ratio (SNR) 
is sufficiently high. In addition, the MAC protocol in these works was the standard window-based CSMA MAC protocol, which is known to be outperformed by the p-persistent CSMA 
MAC protocol \cite{Cali00}.

Optimal sensing and access design for CRNs were designed by using optimal stopping
theory in \cite{jia08}. In \cite{Sala10}, a multi-channel MAC protocol was proposed considering the distance
among users so that white spaces can be efficiently exploited while satisfactorily protecting primary users (PUs).
Different power and spectrum allocation algorithms were devised to maximize the secondary network throughput in
\cite{taosu10}-\cite{zhang113}. Optimization of spectrum sensing and access in which either cellular or TV bands can be
employed was performed in \cite{choi11}. These existing works either assumed perfect spectrum sensing or did not consider the
cooperative spectrum sensing in their design and analysis. 

Cooperative spectrum sensing has been proposed to improve the sensing performance where several SUs collaborate with each other to identify 
spectrum holes  \cite{Gan07}-\cite{Seu10}  and \cite{Chaud12}. 
In a typical cooperative sensing scheme, each SU performs sensing independently and then sends its sensing result to a central controller (e.g., an access point (AP)). 
Here, various aggregation rules can be employed to combine these sensing results at the central controller to decide whether or not a particular spectrum band is available for secondary access. 
In \cite{Chaud12}, the authors studied the performance of hard decisions and soft decisions at a fusion center. 
They also investigated the impact of reporting channel errors on the cooperative sensing performance.
Recently, the authors of \cite{Lee13} proposed a novel cooperative spectrum sensing scheme using hard decision combining considering feedback errors.
In \cite{Peh09}-\cite{Wei09}, optimization of cooperative sensing under the a-out-of-b rule was studied. 
In \cite{Wei11}, the game-theoretic based method was proposed for cooperative spectrum sensing.
In \cite{Seu10}, the authors investigated the multi-channel scenario where the AP collects statistics from SUs to decide whether it should stop at the current time slot.  
In \cite{Male11, Male13}, two different optimization problems for cooperative sensing were studied.
The first one focuses on throughput maximization where the objective is the probability of false alarm.
The second one attempts to perform interference management where the objective is the probability of detection.
These existing works focused on designing and optimizing parameters for the cooperative spectrum sensing algorithm; however,
they did not consider spectrum access issues. Furthermore, either the single channel setting or homogeneous network scenario (i.e., SUs experience
the same channel condition and spectrum statistics for different channels) was assumed in these works. 

In \cite{zhang11} and \cite{Park11}, the authors conducted design and analysis for cooperative spectrum sensing and MAC protocol design for cognitive radios where parallel spectrum sensing on different channels was assumed to be performed by multiple spectrum sensors at each SU. 
In CRNs with parallel-sensing, there is no need to optimize spectrum sensing sets for SUs. 
These works again considered the homogeneous network and each SU simply 
senses all channels. To the best of our knowledge, existing cooperative spectrum sensing schemes rely
on a central controller to aggregate sensing results for white space detection (i.e., centralized design). In addition, homogeneous
 environments and parallel sensing have been commonly assumed in the literature, which would not be very realistic.

In this work, we consider a general SDCSS and access framework under the heterogeneous environment where statistics of wireless channels, and spectrum holes can be arbitrary and there is no central controller to collect sensing results and make spectrum status decisions. 
In addition, we assume that each SU is equipped with only one spectrum sensor so that SUs have to sense channels sequentially. 
This assumption would be applied to real-world hardware-constrained cognitive radios. 
The considered SDCSS scheme requires SUs to perform sensing on their assigned sets of channels and
then exchange spectrum sensing results with other SUs, which can be subject to errors. 
After the sensing and reporting phases, SUs employ the $p$-persistent CSMA MAC protocol \cite{Cali00} to access one available channel.
In this MAC protocol, parameter $p$ denotes the access probability to the chosen channel if the carrier sensing indicates an available
 channel (i.e., no other SUs transmit on the chosen channel).
It is of interest to determine the access parameter $p $ that can mitigate the collisions and hence enhance the system throughput \cite{Cali00}.
Also, optimization of the spectrum sensing set for each SU (i.e., the set of channels sensed by the SU)  is very critical to achieve good system throughput.
Moreover, analysis and optimization of the joint spectrum sensing and access design become
much more challenging in the heterogeneous environment, which, however, can significantly improve the system performance.
Our current paper aims to resolve these challenges whose contributions can be summarized as follows:
 
\begin{itemize}

\item We propose the distributed $p$-persistent CSMA protocol incorporating SDCSS for multi-channel CRNs. 
Then we analyze the saturation throughput and optimize the spectrum sensing time and access parameters to achieve maximum throughput 
for a given allocation of channel sensing sets. This analysis and optimization are performed in the general heterogeneous scenario assuming that spectrum sensing sets 
for SUs have been predetermined.

\item We study the channel sensing set optimization (i.e., channel assignment) for throughput maximization and devise both
exhaustive search and low-complexity greedy algorithms to solve the underlying NP-hard optimization problem.
Specifically, an efficient solution for the considered problem would only allocate a subset of ``good'' SUs to sense each channel 
so that accurate sensing can be achieved with minimal sensing time.
We also analyze the complexity of the brute-force search and the greedy algorithms.

\item We extend the design and analysis to consider reporting errors as SUs exchange their spectrum sensing results.
In particular, we describe cooperative spectrum sensing model, derive the saturation throughput
considering reporting errors. Moreover, we discuss how the proposed algorithms to optimize the sensing/access parameters and sensing sets
can be adapted to consider reporting errors. Again, all the analysis is performed for the heterogeneous environment.

\item We present numerical results to illustrate the impacts of different parameters on the secondary throughput performance and demonstrate the 
significant throughput gain due to the optimization of different parameters in the proposed framework.

\end{itemize}

The remaining of this paper is organized as follows. Section ~\ref{SystemModel} describes system and sensing models. 
MAC protocol design, throughput analysis, and optimization are performed in Section ~\ref{CPCSMA} assuming no reporting errors. 
Section ~\ref{Exten} provides further extension for the analysis and optimization considering reporting errors.
Section ~\ref{Results} presents numerical results followed by concluding remarks in Section ~\ref{conclusion}.
The summary of key variables in the paper is given in Table~IV.
 
%\vspace{10pt}
\section{System Model and Spectrum Sensing Design}
\label{SystemModel}

In this section, we describe the system model and spectrum sensing design for the multi-channel CRNs.
Specifically, sensing performances in terms of detection and false alarm probabilities are presented.

\subsection{System Model}
\label{System}

 %Fig. 1
\begin{figure}[!t] 
\centering
\includegraphics[width=70mm]{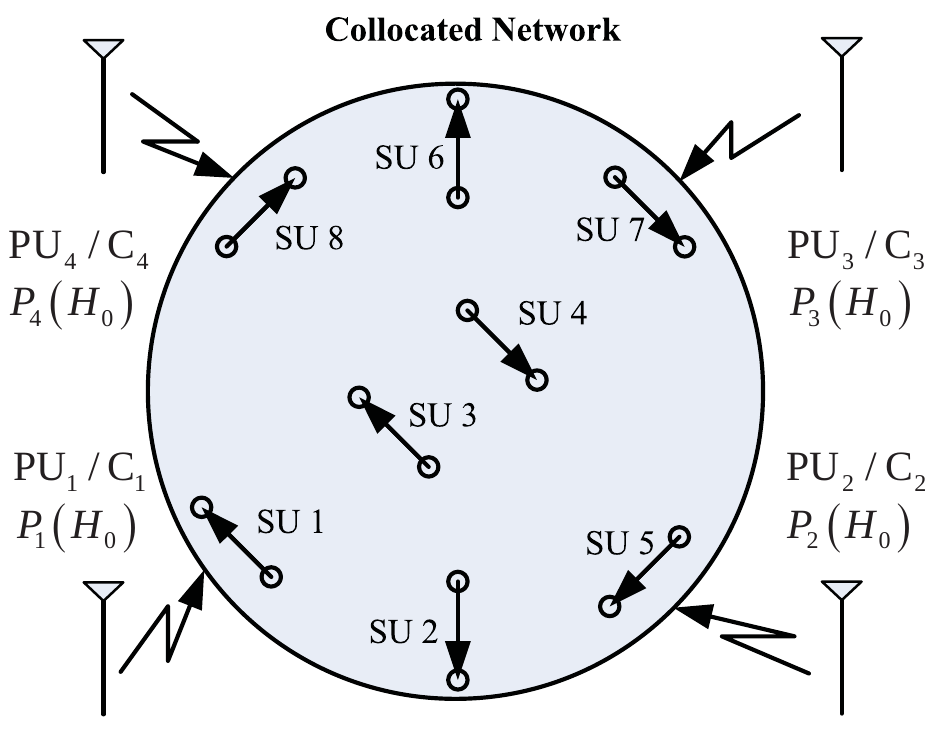}
\caption{Considered network and spectrum sharing model (PU: primary user, SU: secondary user, and $C_i $ is the channel $i $ corresponding to $\text{PU}_i $)}
\label{Fig1}
\end{figure}

We consider a network setting where $N$ pairs of SUs opportunistically exploit white spaces
 in $M$ channels for data transmission. For simplicity, we refer to pair $i$ of SUs simply as SU $i$. 
We assume that each SU can exploit only one available channel for transmission (i.e., SUs are equipped with 
narrow-band radios). We will design a synchronized MAC protocol integrating SDCSS for channel access. We assume 
that each channel is either in the idle or busy state for each predetermined periodic interval, which is referred to as
a cycle in this paper.

We further assume that each pair of SUs can overhear transmissions from other pairs of SUs (i.e., collocated networks). 
There are $M$ PUs each of which may or may not use one corresponding channel for its data transmission
in each cycle. In addition, it is assumed that transmission from any pair of SUs on a particular channel will affect the primary receiver
which receives data on that channel. The network setting under investigation is shown in Fig.~\ref{Fig1} where $C_i$
denotes channel $i$ that belongs to PU $i$.
 
\subsection{Semi-Distributed Cooperative Spectrum Sensing}
\label{Ss}

We assume that each SU $i$ is assigned a set of channels $\mathcal{S}_i $ where
it senses all channels in this assigned set at beginning of each cycle in a sequential manner (i.e., sense one-by-one).
Optimization of such channel assignment will be considered in the next section.
Upon completing the channel sensing, each SU $i$ exchanges the sensing results (i.e., idle/busy status
of all channels in $\mathcal{S}_i $) with other SUs for further processing. 
Here, the channel status of each channel can be represented by one bit (e.g., 1 for idle and 0 for busy status).
Upon collecting sensing results, each SU will decide idle/busy status for all channels. Then, 
SUs are assumed to employ a distributed MAC protocol to perform access resolution so that only
the winning SUs on each channel are allowed to transmit data. The detailed MAC protocol design will be presented later.

Let $\mathcal{H}_0$ and $\mathcal{H}_1$ denote the events that a particular PU is idle and active on its corresponding channel in any cycle, respectively. 
In addition, let $\mathcal{P}_j \left( \mathcal{H}_0 \right)$ and  $\mathcal{P}_j \left( \mathcal{H}_1 \right) = 1 - 
\mathcal{P}_j \left( \mathcal{H}_0 \right)$ be the probabilities that channel $j$ is available and not available for secondary access, respectively.
We assume that SUs employ an energy detection sensing scheme and let $f_s$ be the sampling frequency used in the
sensing period for all SUs. There are two important performance measures,
which are used to quantify the sensing performance, namely detection and false alarm probabilities. In particular, a
detection event occurs when a SU successfully senses a busy channel and false alarm
represents the situation when a spectrum sensor returns a busy status for an idle channel (i.e., the transmission opportunity
is overlooked). 
 
Assume that transmission signals from PUs are complex-valued PSK signals while the noise at the SUs is independent and identically distributed circularly 
symmetric complex Gaussian $\mathcal{CN}\left( {0,{N_0}} \right)$ \cite{R1}. Then, the
detection and false alarm probabilities experienced by SU $i$ for the channel $j$ can be calculated as \cite{R1}
\beqn
\label{eq1}
\mathcal{P}_d^{ij}\left( \varepsilon ^{ij} ,\tau^{ij}  \right) = \mathcal{Q}\left( \left( \frac{\varepsilon ^{ij} }{N_0} - \gamma ^{ij}  - 1 \right)\sqrt {\frac{\tau^{ij} f_s}{2\gamma ^{ij}  + 1}}  \right), 
\eeqn
\beqn
 \mathcal{P}_f^{ij}\left( \varepsilon ^{ij} ,\tau^{ij}  \right) = \mathcal{Q}\left( \left( \frac{\varepsilon ^{ij} }{N_0} - 1 \right)\sqrt {\tau^{ij} f_s}  \right) \hspace{1.5cm} \nonumber \\ 
 = \mathcal{Q}\left( \sqrt {2\gamma ^{ij}  + 1} \mathcal{Q}^{ - 1}\left( \mathcal{P}_d^{ij}\left(  \varepsilon ^{ij} ,\tau^{ij}   \right) \right)+\sqrt {\tau^{ij} f_s} \gamma ^{ij}  \right),  \label{eq2}
\eeqn
where $i \in \left[ {1,N} \right]$ is the SU index, $j \in \left[ {1,M} \right]$ is the channel index, ${\varepsilon ^{ij}} $ is the detection threshold for 
the energy detector, ${\gamma ^{ij}} $ is the signal-to-noise ratio (SNR) of the PU's signal at the SU, $f_s$ is the sampling frequency, $N_0$ is the noise power, 
$\tau^{ij}$ is the sensing time of SU $i$ on channel $j$, and $\mathcal{Q}\left( . \right)$ is defined as 
$\mathcal{Q}\left( x \right) = \left( {1/\sqrt {2\pi } } \right)\int_x^\infty  {\exp \left( { - {t^2}/2} \right)dt}$. 

We assume that a general cooperative sensing scheme, namely $a$-out-of-$b$ rule, is employed by each SU to determine the idle/busy status
of each channel based on reported sensing results from other SUs. Under this scheme, an SU will declare that a channel is busy
if $a$ or more messages out of $b$ sensing messages report that the underlying channel is busy. 
The a-out-of-b rule covers different rules including OR, AND and majority rules as special cases. In particular, $a=1$ corresponds to the OR rule; 
if $a=b$ then it is the AND rule; and the majority rule has $a=\left\lceil b/2\right\rceil$. 

\begin{figure}[!t]
\centering
\includegraphics[width=40mm]{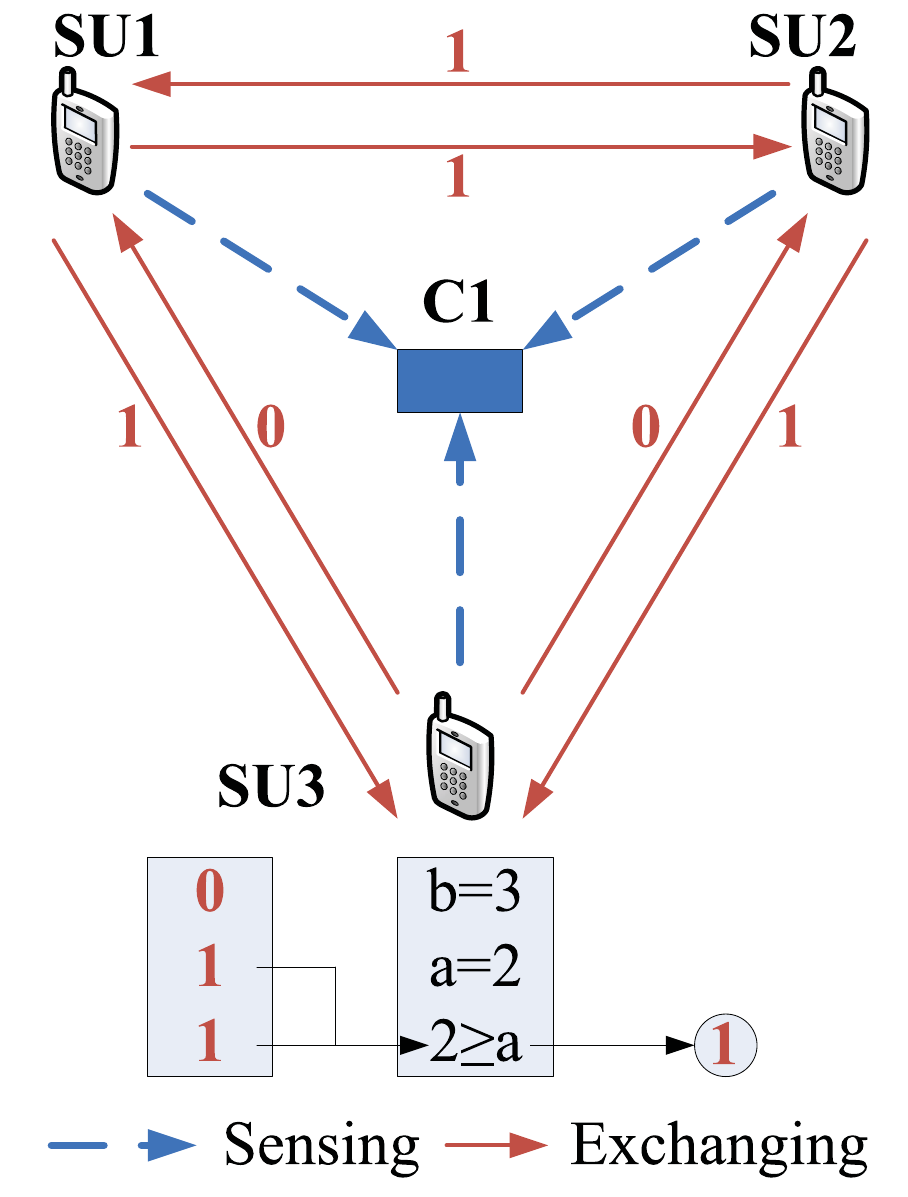}
\caption{Example for SDCSS on 1 channel.}
\label{DCSS_eg}
\end{figure}

To illustrate the operations of the $a$-out-of-$b$ rule, let us consider a simple example shown in Fig.~\ref{DCSS_eg}.
Here, we assume that 3 SUs collaborate to sense channel one with $a = 2 $ and $b = 3 $.
After sensing channel one, all SUs exchange their sensing outcomes. 
SU3 receives the reporting results comprising two ``1'' and one ``0'' where ``1'' means that the channel is busy and ``0''  means 
channel is idle. Because the total number of ``1s'' is two which is larger than or equal to $a=2$, SU3 outputs the ``1'' in the final sensing result,
namely the channel is busy.

Let us consider a particular channel $j$. Let $\mathcal{S}_j^U$ denote the set of SUs that sense channel $j$, $b_j=\left|\mathcal{S}_j^U\right|$
be the number of SUs sensing channel $j$, and $a_j $ be the number of messages indicating that the underlying channel is busy. 
Then, the final decision on the spectrum status of channel $j$ under the $a$-out-of-$b$ rule has detection and false alarm probabilities that can be written as \cite{Wei11}
\beqn
\mathcal{P}_u^j\left( {\vec \varepsilon ^j}, {\vec \tau^j} , a_j  \right) = \sum_{l=a_j}^{b_j} \sum_{k=1}^{C_{b_j}^l} \prod_{i_1 \in \Phi^k_l} \mathcal{P}_u^{i_1j} \prod_{i_2 \in \mathcal{S}_j^{U} \backslash \Phi^k_l} \mathcal{\bar P}_u^{i_2j} \label{eq1_css_1},
\eeqn
where $u$ represents $d$ or $f$ as we calculate the probability of detection $\mathcal{P}_d^j$ or false alarm $\mathcal{P}_f^j$, respectively; $\mathcal{\bar P}$ is defined as
$\mathcal{\bar P} = 1-\mathcal{P}$; $\Phi^k_l$ in (\ref{eq1_css_1}) denotes a particular set with $l$ SUs whose sensing outcomes suggest that channel $j$ is busy given that 
this channel is indeed busy and idle as $u$ represents $d$ and $f$, respectively. Here, we generate all possible combinations of $\Phi^k_l$ where there are indeed
$C_{b_j}^l $ combinations. Also, ${\vec \varepsilon ^j} = \left\{\varepsilon ^{ij} \right\}$, ${\vec \tau^j} = \left\{\tau^{ij}\right\}$, $i \in \mathcal{S}_j^U$ represent the
 set of detection thresholds and sensing times, respectively. 
For brevity, $\mathcal{P}_d^j\left({\vec \varepsilon ^j}, {\vec \tau^j}, a_j  \right) $ and  $\mathcal{P}_f^j\left( {\vec \varepsilon ^j}, {\vec \tau^j}, a_j  \right)$ are sometimes written as $\mathcal{P}_d^j$ and $\mathcal{P}_f^j$ in the following.

Each SU exchanges the sensing results on its assigned channels with other SUs over a control channel, which is assumed to be always available (e.g., it is owned by the secondary network). 
To avoid collisions among these  message exchanges, we assume that there are $N$ reporting time slots for $N$ SUs each of which
has length equal to $t_r$. Hence, the total time for exchanging sensing results among SUs is $Nt_r$. 
Note that the set of channels assigned to SU $i$ for sensing, namely $\mathcal{S}_i $, is a subset of all channels and these sets can be different for different SUs. 
An example of channel assignment (i.e., channel sensing sets) is presented in Table \ref{table}. In this table, SU 4 is not assigned any channel. Hence, this 
SU must rely on the sensing results of other SUs to determine the spectrum status. 

\begin{table} 
\centering
\caption{Channel Assignment Example for SUs (x denotes an assignment)}
\label{table}
\begin{tabular}{|c|c|c|c|c|c|c|}
\cline{3-7} 
\multicolumn{2}{c|}{} & \multicolumn{5}{c|}{\textbf{Channel}}\tabularnewline
\cline{3-7} 
\multicolumn{2}{c|}{} & 1 & 2 & 3 & 4 & 5\tabularnewline
\hline 
 & 1 & x &  & x & x & \tabularnewline
\cline{2-7} 
 & 2 & x & x &  &  & \tabularnewline
\cline{2-7} 
\textbf{SU} & 3 &  & x & x & x & \tabularnewline
\cline{2-7} 
 & 4 &  &  &  &  & \tabularnewline
\cline{2-7} 
 & 5 & x &  &  &  & x\tabularnewline
\hline
\end{tabular}
\end{table}

%\noindent
\textit{Remark 1:} In practice, the idle/busy status of primary system on a particular channel can be arbitrary and would not be synchronized with 
the operations of the SUs (i.e., the idle/busy status of any channel can change in the middle of a cycle). 
Hence, to strictly protect the PUs, SUs should continuously scan the spectrum of interest and evacuate from an exploited channel as soon as
the PU changes from an idle to a busy state. However, this continuous spectrum monitoring would be very costly to implement 
since each SU should be equipped with two half-duplex transceivers to perform spectrum sensing and access at the same time.
A more efficient protection method for PUs is to perform periodic spectrum sensing where SUs perform spectrum sensing at the beginning of each fixed-length
 interval and exploits available frequency bands for data transmission during the remaining time of the interval.
In this paper, we assume that the idle/busy status of each channel remains the same in each cycle, which enables us to analyze the system throughput. 
In general, imposing this assumption would not sacrifice the accuracy of our throughput analysis if PUs maintain their idle/busy status for a 
sufficiently long time. This is actually the case for many practical scenarios such as in the TV bands, as reported by several recent studies [34]. 
In addition, our MAC protocol that is developed under this assumption would result in very few collisions with PUs because the cycle time is quite small compared 
to the typical intervals over which the active/idle statuses of PUs change.

\vspace{10pt}
\section{Performance Analysis and Optimization for Cognitive MAC Protocol}
\label{CPCSMA}

We present the cognitive MAC protocol design, performance analysis, and optimization for the multi-channel CRNs in this section.

\subsection{Cognitive MAC Protocol Design}
\label{MACDesign}

\begin{figure*}%[!t]
\centering
\includegraphics[width=170mm]{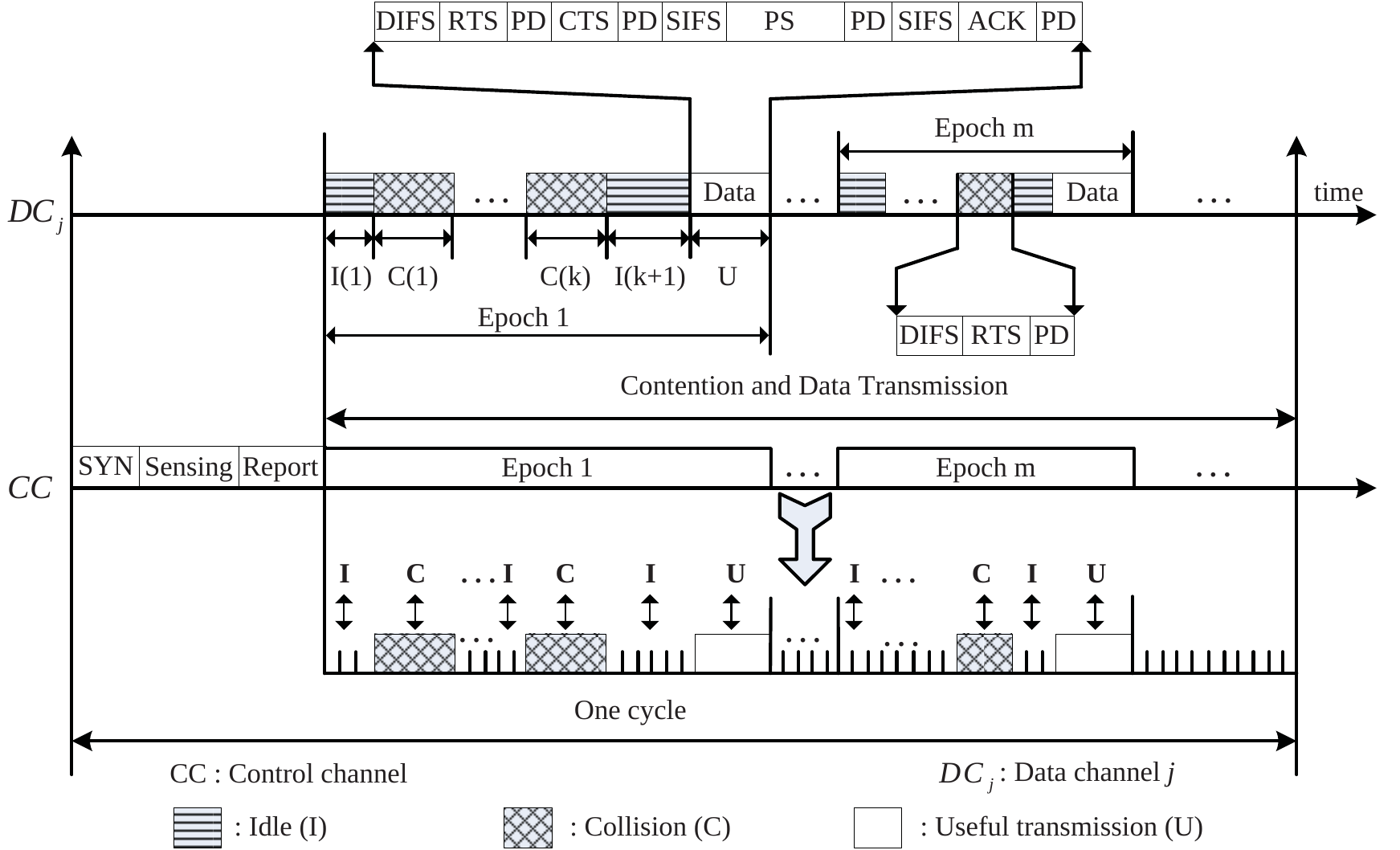}
\caption{Timing diagram of cognitive $p$-persistent CSMA protocol for one specific channel $j$.}
\label{Fig3_1}
\end{figure*}

We assume that time is divided into fixed-size cycles and it is assumed that SUs can perfectly synchronize with each other (i.e., there is no synchronization error) \cite{Konda08}.
We propose a synchronized multi-channel MAC protocol for dynamic spectrum sharing as follows.
The MAC protocol has four phases in each cycle as illustrated in Fig.~\ref{Fig3_1}.
The beacon signal is sent on the control channel to achieve synchronization in the first phase \cite{Konda08} which is presented in the simple manner as follows.
At the beginning of this phase, each SU senses the beacon signal from the volunteered synchronized SU which is the first SU sending the beacon. 
If an SU does not receive any beacon, it selects itself as the volunteered SU and sends out the beacon for synchronization.
In the second phase, namely the sensing phase of length $\tau$, all SUs simultaneously perform spectrum sensing on their assigned channels. 
Here, we have $\tau = \max_{i} \tau^i$, where $\tau^i = \sum_{j \in \mathcal{S}_i} \tau^{ij}$ is total sensing time of SU $i$, $\tau^{ij}$ is the sensing time of SU $i$ on channel $j$, and $\mathcal{S}_i$ is the set of channels assigned for SU $i$. 
We assume that one separate channel is assigned as a control channel which is used to exchange sensing results for reporting as well as broadcast a beacon signal for synchronization.
This control channel is assumed to be always available (e.g., it is owned by the secondary network).
In the third phase, all SUs exchange their sensing results with each other via the control channel.
Based on these received sensing results, each SU employs SDCSS techniques to decide the channel status of all channels and hence has a set of available channels.
Then each SU transmitter will choose one available channel randomly (which is used for contention and data transmission) and inform it to the corresponding SU receiver via the control channel.

In the fourth phase, SUs will participate in contention and data transmission on their chosen channels. 
We assume that the length of each cycle is sufficiently large so that SUs can transmit several packets during this data contention and transmission phase. 
In particular, we employ the $p$-persistent CSMA principle \cite{Cali00} to devise our cognitive MAC protocol.
In this protocol, each SU attempts to transmit on the chosen channel with a probability of $p$ if it senses an available channel (i.e., no other SUs transmit data on its chosen channel). 
In case the SU decides not to transmit (with probability of $1-p$), it will sense the channel and attempt to transmit again in the next slot with probability $p$.
If there is a collision, the SU will wait until the channel is available and attempt to transmit with probability $p$ as before.

The standard 4-way handshake with RTS/CTS (request-to-send/clear-to-send) \cite{R3} will be employed to reserve a channel for data transmission. 
So the SU choosing to transmit on each available channel  exchanges RTS/CTS messages before transmitting its actual data packet. 
An acknowledgment (ACK) from the receiver is transmitted to the transmitter for successful reception of any packet.
The detailed timing diagram of this MAC protocol is presented in Fig.~\ref{Fig3_1}. 

%\noindent
\textit{Remark 2:} For simplicity, we consider the fixed control channel in our design.
However, extensions to consider dynamic control channel selections to avoid the congestion can be adopted in our proposed framework.
More information on these designs can be found in \cite{Mo08}.

\vspace{10pt}
\label{TputPCSMA}
\subsection{Saturation Throughput Analysis}

In this section, we analyze the saturation throughput of the proposed cognitive $p$-persistent CSMA protocol
assuming that there are no reporting errors in exchanging the spectrum sensing results among SUs.
Because there are no reporting errors, all SUs acquire the same sensing results for each channel, which implies that they make the same final sensing decisions
since the same a-out-b aggregation rule is employed for each channel. In the analysis, transmission time is counted in terms of 
contention time slot, which is assumed to be $v$ seconds. Each data packet is assumed to be of fixed size of $PS$ time slots. Detailed timing diagram 
of the $p$-persistent CSMA MAC protocol is illustrated in Fig.~\ref{Fig3_1}.

Any particular channel alternates between idle and busy periods from the viewpoint of the secondary system  where each busy period
 corresponds to either a collision or a successful transmission. We use the term ``epoch'' to refer to the interval between two consecutive successful transmissions. 
This means an epoch starts with an idle period followed by some alternating collision periods and idle periods before ending with a successful transmission period. 
Note that an idle period corresponds to the interval between two consecutive packet transmissions (collisions or successful transmissions).

Recall that each SU chooses one available channel randomly for contention and transmission according to the final cooperative sensing outcome.
We assume that upon choosing a channel, an SU keeps contending and accessing this channel until the end of the current cycle.
In the case of missed detection (i.e., the PU is using the underlying channel but the sensing outcome suggests that the channel is available), there will be 
collisions between SUs and the PU. Therefore,  RTS and CTS exchanges will not be successful in this case even though SUs cannot 
differentiate whether they collide with other SUs or the PU. 
Note that channel accesses of SUs due to missed detections do not contribute to the secondary system throughput.

To calculate the throughput for the secondary network, we have to consider all scenarios of idle/busy statuses of all channels and possible
mis-detection and false alarm events for each particular scenario. Specifically,
the normalized throughput per one channel achieved by our proposed MAC protocol, $\mathcal{NT} \left( \left\{\tau^{ij}\right\}, \left\{a_j\right\}, p, \left\{\mathcal{S}_i\right\} \right) $ can be written as 
\beqn
\mathcal{NT} = \sum_{k_0=1}^M \sum_{l_0=1}^{C_M^{k_0}} \prod_{j_1 \in \Psi_{k_0}^{l_0}} \mathcal{P}_{j_1} \left(\mathcal{H}_0\right) \prod_{j_2 \in \mathcal{S} \backslash \Psi_{k_0}^{l_0}} \mathcal{P}_{j_2} \left(\mathcal{H}_1\right) \times \label{NTSC_11}\\
\sum_{k_1=1}^{k_0} \sum_{l_1=1}^{C_{k_0}^{k_1}} \prod_{j_3 \in \Theta_{k_1}^{l_1}} \mathcal{\bar P}_f^{j_3} \prod_{j_4 \in \Psi_{k_0}^{l_0} \backslash \Theta_{k_1}^{l_1}} \mathcal{P}_f^{j_4}  \times \label{NTSC_21}\\
\sum_{k_2 = 0}^{M-k_0} \sum_{l_2=1}^{C_{M-k_0}^{k_2}} \prod_{j_5 \in \Omega_{k_2}^{l_2}} \mathcal{\bar P}_d^{j_5} \prod_{j_6 \in \mathcal{S} \backslash \Psi_{k_0}^{l_0} \backslash \Omega_{k_2}^{l_2}} \mathcal{P}_d^{j_6} \times  \label{NTSC_31}\\
\mathcal{T}_p^{\sf ne} \left(\tau,\left\{a_j\right\},p  \right). \label{NTSC_41}
\eeqn
The quantity (\ref{NTSC_11}) represents the probability that there are $k_0$ available channels, which may or may not  be correctly determined by the SDCSS. 
Here, $\Psi_{k_0}^{l_0} $ denotes a particular set of $k_0 $ available channels out of $M$ channels whose index is $l_0$. 
In addition, the quantity (\ref{NTSC_21}) describes the probability that the SDCSS indicates $k_1 $ available channels whereas the remaining available channels are 
overlooked due to sensing errors where $\Theta_{k_1}^{l_1} $ denotes the $l_1 $-th set with $k_1 $ available channels. 
For the quantity in (\ref{NTSC_31}), $k_2$ represents the number of channels that are not available but the sensing outcomes indicate that they are available 
(i.e., due to misdetection) where $\Omega_{k_2}^{l_2}$ denotes the $l_2$-th set with $k_2$ mis-detected channels. 
The quantity in (\ref{NTSC_31}) describes the probability that the sensing outcomes due to SUs incorrectly indicates $k_2$ available channels. Finally,
$\mathcal{T}_p^{\sf ne} \left(\tau,\left\{a_j\right\},p  \right)$ in (\ref{NTSC_41}) denotes the conditional throughput for a particular realization of sensing
outcomes corresponding to two sets $\Theta_{k_1}^{l_1} $ and $\Omega_{k_2}^{l_2}$. 

Therefore, we have to derive the conditional throughput $\mathcal{T}_p^{\sf ne} \left(\tau,\left\{a_j\right\},p \right)$ to complete the throughput analysis, which
is pursued in the following. Since each SU randomly chooses one available channel according to the SDCSS for contention and access, the number of SUs
actually choosing a particular available channel is a random number. In addition, the SDCSS suggests that channels in $\Theta_{k_1}^{l_1} \cup \Omega_{k_2}^{l_2}$
are available for secondary access but only channels in $\Theta_{k_1}^{l_1}$ are indeed available and can contribute to the secondary throughput 
 (channels in $\Omega_{k_2}^{l_2}$ are misdetected by SUs). Let $\left\{n_{j}\right\} = \left\{n_1, n_2, \ldots, n_{k_e}\right\}$
be the vector describing how SUs choose channels for access where  $k_e = \left|\Theta_{k_1}^{l_1} \cup \Omega_{k_2}^{l_2}\right|$ and $n_{j}$
denotes the number of SUs choosing channel $j$ for access. Therefore, the conditional throughput  
$\mathcal{T}_p^{\sf ne} \left(\tau,\left\{a_j\right\},p \right)$ can be calculated as follows:
\beqn
\label{T_p_cal}
\mathcal{T}_p^{\sf ne}  \left(\tau,\left\{a_j\right\},p \right) = \sum_{\left\{n_{j}\right\}: \: \sum _{j \in \Theta_{k_1}^{l_1} \cup \Omega_{k_2}^{l_2}} n_{j}=N} 
\mathcal{P}\left(\left\{n_{j}\right\}\right) \times \label{T_p_cal_11} \\
\sum_{j_2 \in \Theta_{k_1}^{l_1}} \frac{1}{M}  \mathcal{T}_{j_2}^{\sf ne} \left(\tau,\left\{a_{j_2}\right\},p \left|n=n_{j_2}\right.\right)  
\mathcal{I} \left(n_{j_2}>0\right) \label{T_p_cal_22}, \label{thputpnv}
\eeqn
where $\mathcal{P}\left(\left\{n_{j}\right\}\right)$ in (\ref{T_p_cal_11}) represents the probability that the channel access vector $\left\{n_{j}\right\}$ is realized (each channel $j$ where $j \in \Theta_{k_1}^{l_1} \cup \Omega_{k_2}^{l_2}$ is selected by $n_{j} $ SUs). 
The sum in (\ref{thputpnv}) describes the normalized throughput per channel due to a particular realization of the access vector $\left\{n_{j}\right\}$. 
Therefore, it is equal to the total throughput achieved by all available channels (in the set $\Theta_{k_1}^{l_1}$) divided by the total number of channels $M$.
Here, $\mathcal{T}_{j_2}^{\sf ne} \left(\tau,\left\{a_{j_2}\right\},p \left|n=n_{j_2}\right.\right) $ denotes the conditional throughput achieved by a particular channel $j_2$ when there are $n_{j_2}$ contending on this channel and $\mathcal{I} \left(n_{j_2}>0\right)$ represents the indicator function, which is equal to zero if $n_{j_2}=0$ (i.e., no SU chooses channel $j_2$) and equal to one, otherwise. 
Note that the access of channels in the set $\Omega_{k_2}^{l_2}$ due to missed detection does not contribute to the system throughput, which explains why we do not include these channels in the sum in (\ref{thputpnv}).

Therefore,  we need to drive $\mathcal{P}\left(\left\{n_{j}\right\}\right)$ and $\mathcal{T}_{j_2}^{\sf ne} \left(\tau,\left\{a_{j_2}\right\},p \left|n=n_{j_2}\right.\right)$ to determine the normalized throughput. Note that the sensing outcome due to the SDCSS is the same for all SUs and each SU chooses one channel in the set of
 $k_e = \left|\Theta_{k_1}^{l_1} \cup \Omega_{k_2}^{l_2}\right|$ channels randomly. Therefore, the probability $\mathcal{P}\left(\left\{n_{j}\right\}\right)$  can be
 calculated as follows:
\beqn
\mathcal{P}\left(\left\{n_{j}\right\}\right) &=& \left( {\begin{array}{*{20}{c}}
   N  \\
   {\left\{n_{j}\right\}}  \\
\end{array}} \right) \left(\frac{1}{k_e}\right)^{\sum _{j \in \Theta_{k_1}^{l_1} \cup \Omega_{k_2}^{l_2}} n_{j}} \\
&=& \left( {\begin{array}{*{20}{c}}
   N  \\
   {\left\{n_{j}\right\}}  \\
\end{array}} \right) \left(\frac{1}{k_e}\right)^N,
\eeqn
where $\left( {\begin{array}{*{20}{c}}
   N  \\
   {\left\{n_{j}\right\}}  \\
\end{array}} \right)$ is the multinomial coefficient which is defined as $ \left( {\begin{array}{*{20}{c}}
   N  \\
   {\left\{n_{j}\right\}}  \\
\end{array}} \right) = \left( {\begin{array}{*{20}{c}}
   N  \\
   {n_1, n_2, \ldots, n_k}  \\
\end{array}} \right) = \frac{N!}{n_1! n_2! \ldots n_k!}$.

The calculation of the conditional throughput $\mathcal{T}_{j_2}^{\sf ne} \left(\tau,\left\{a_{j_2}\right\},p \left|n=n_{j_2}\right.\right)$ must account for 
the overhead due to spectrum sensing and exchanges of sensing results among SUs. Let us define $T_R = Nt_r $ where $t_r$ is the report time from each SU to 
all the other SUs; $\tau = \max_{i} \tau^i$ is the total the sensing time; ${\bar T}^{j_2}_{\sf cont}$ is the average total  time due to contention, collisions, 
and RTS/CTS exchanges before a successful packet transmission; $T_S$ is the total time for transmissions of data packet, ACK control packet, and overhead between these data
and ACK packets. Then, the conditional throughput $\mathcal{T}_{j_2}^{\sf ne} \left(\tau,\left\{a_{j_2}\right\},p \left|n=n_{j_2}\right.\right)$ can be written as 
\beqn
\label{con_T}
\mathcal{T}_{j_2}^{\sf ne} \left(\tau,\left\{a_{j_2}\right\},p \left|n=n_{j_2}\right.\right) = \left\lfloor \frac{T-\tau-T_R}{{\bar T}^{j_2}_{\sf cont} + T_S}\right\rfloor \frac{T_S}{T},
\eeqn
where $\left\lfloor  .  \right\rfloor $ denotes the floor function and recall that $T$ is the duration of a cycle. Note
that $\left\lfloor \frac{T - \tau - T_R}{{\bar T}^{j_2}_{\sf cont} + T_S} \right\rfloor$ denotes the average number of successfully transmitted packets
in one particular cycle excluding the sensing and reporting phases. Here, we omit the length of the synchronization phase, which is assumed to be negligible.

To calculate ${\bar T}^{j_2}_{\sf cont} $, we define some further parameters as follows.
Let denote $T_C$ as the duration of the collision; 
${\bar T}_S$ is the required time for successful RTS/CTS transmission. These quantities can be calculated under the 4-way handshake mechanism as \cite{Cali00}
\beqn
\label{TCTSTI}
\left\{ {\begin{array}{*{20}{c}}
   T_S = PS + 2SIFS + 2PD + ACK   \hfill\\
   {\bar T}_S = DIFS + RTS + CTS + 2PD \hfill  \\
   T_C = RTS + DIFS + PD \hfill  \\
\end{array}} \right.,
\eeqn
where $PS$ is the packet size, $ACK$ is the length of an ACK packet, $SIFS$ is the length of a short interframe space, $DIFS$ is the length of a distributed interframe space, 
$PD$ is the propagation delay where $PD$ is usually very small compared to the slot size $v$. 

Let $T_I^{i,j_2}$ be the $i$-th idle duration between two consecutive RTS/CTS transmissions
 (they can be collisions or successes) on a particular channel $j_2$. Then, $T_I^{i,j_2}$ can be calculated based on its probability mass function (pmf),
 which is derived in the following. Recall that all quantities are defined in terms of number of time slots. Now, suppose there are  $n_{j_2}$ SUs
choosing channel $j_2$, let $\mathcal{P}_S^{j_2}$, $\mathcal{P}_C^{j_2}$ and $\mathcal{P}_I^{j_2}$ be the probabilities of a generic slot corresponding to a successful transmission, a collision and an idle slot, respectively. These quantities are calculated as follows
\beqn
\mathcal{P}_S^{j_2} = n_{j_2}p\left(1-p\right)^{n_{j_2}-1} \\
\mathcal{P}_I^{j_2} = \left(1-p\right)^{n_{j_2}} \\
\mathcal{P}_C^{j_2} = 1-\mathcal{P}_S^{j_2}-\mathcal{P}_I^{j_2},
\eeqn
where $p$ is the transmission probability of an SU in a generic slot. 
Note that ${\bar T}^{j_2}_{\sf cont}$ is a random variable (RV) consisting of several intervals corresponding to idle periods, collisions, and one successful RTS/CTS transmission. 
Hence this quantity for channel $j_2$ can be written as 
\beqn
\label{T_cont}
{\bar T}^{j_2}_{\sf cont} = \sum_{i=1}^{N_c^{j_2}} \left(T_C+ T_I^{i,{j_2}}\right) + T_I^{N_c^{j_2}+1,{j_2}} + {\bar T}_S,
\eeqn
where $N_c^{j_2}$ is the number of collisions before the first successful RTS/CTS exchange. 
Hence it is a geometric RV with parameter $1-\mathcal{P}_C^{j_2}/\mathcal{\bar P}_I^{j_2}$ (where $\mathcal{\bar P}_I^{j_2} = 1 - \mathcal{P}_I^{j_2}$). 
Its pmf can be expressed as
\beqn
\label{N_c_cal}
 f_{X}^{N_c} \left(x\right) = \left(\frac{\mathcal{P}_C^{j_2}}{\mathcal{\bar P}_I^{j_2}}\right)^{x} \left(1-\frac{\mathcal{P}_C^{j_2}}{\mathcal{\bar P}_I^{j_2}}\right), \: x = 0, 1, 2, \ldots
\eeqn
Also, $T_I^{i,j_2}$ represents the number of consecutive idle slots, which is also a geometric RV with parameter $1-\mathcal{P}_I^{j_2}$ with the following pmf
\beqn
\label{T_I_cal}
f_{X}^{I} \left(x\right) = \left(\mathcal{P}_I^{j_2}\right)^{x} \left(1-\mathcal{P}_I^{j_2}\right), \: x = 0, 1, 2, \ldots
\eeqn
Therefore, ${\bar T}^{j_2}_{\sf cont}$  can be written as follows \cite{Cali00}:
\beqn
{\bar T}^{j_2}_{\sf cont}  = {\bar N}_c^{j_2}T_C + {\bar T}_I^{j_2} \left({\bar N}_c^{j_2}+1\right) + {\bar T}_S \label{T_contgeo},
\eeqn
where ${\bar T}_I^{j_2}$ and ${\bar N}_c^{j_2}$ can be calculated as
\beqn
{\bar T}_I^{j_2} &=& \frac{\left(1-p\right)^{n_{j_2}}}{1-\left(1-p\right)^{n_{j_2}}} \\
{\bar N}_c^{j_2} &=& \frac{1-\left(1-p\right)^{n_{j_2}}}{n_{j_2}p\left(1-p\right)^{n_{j_2}-1}}-1. 
\eeqn
These expressions are obtained by using the  pmfs of the corresponding RVs given in (\ref{N_c_cal}) and (\ref{T_I_cal}), respectively \cite{Cali00}.

%\vspace{10pt}
\subsection{Semi-Distributed Cooperative Spectrum Sensing and $p$-persistent CSMA Access Optimization}
\label{OpCSPCSMA}

We determine optimal sensing and access parameters to maximize the normalized throughput for our proposed SDCSS and $p$-persistent CSMA protocol. 
Here, we assume that the sensing sets $\mathcal{S}_j^U$ for different channels $j$ have been given. Optimization of these sensing sets
is considered in the next section. Note that the optimization performed in this paper is different from those in \cite{Le11}, \cite{Le12} because the MAC protocols
and sensing algorithms in the current and previous works are different. The normalized 
throughput optimization problem can be presented as 
\beqn
 \mathop {\max} \limits_{ \left\{\tau^{ij}\right\}, \left\{a_j\right\} , p} \quad \mathcal{NT}_p \left( \left\{\tau^{ij}\right\}, \left\{a_j\right\}, p, \left\{\mathcal{S}_i\right\}  \right)  \hspace{1.4cm} \label{probp1}\\ 
 \mbox{s.t.}\,\,\,\, \mathcal{P}_d^j \left( {\vec \varepsilon ^j}, {\vec \tau^j}, a_j  \right) \geq \mathcal{\widehat{P}}_d ^j, \: j \in \left[1, M\right] \hspace{1cm} \label{probp2}\\
 \quad \quad 0 < \tau^{ij}  \le {T},  \: 0 \leq p \leq 1, \hspace{1cm} \label{probp3}
\eeqn
where $\mathcal{P}_d^j$ is the detection probability for channel $j$; $\mathcal{\widehat{P}}_d ^j$ denotes the target detection probability;
$\vec \varepsilon ^j$ and $\vec \tau^j$ represent the vectors of detection thresholds and sensing times on channel $j$, respectively;
$a_j$ describes the parameter of the $a_j$-out-of-$b_j$ aggregation rule for SDCSS on channel $j$ with $b_j = |\mathcal{S}_j^U|$
where recall that $\mathcal{S}_j^U$ is the set of SUs sensing channel $j$. The optimization variables for this problem are
sensing times $\tau^{ij}$ and parameters $a_j$ of the sensing aggregation rule, and transmission probability $p$ of the MAC protocol.

It was shown in \cite{R1} that the constraints on detection probability should be met with equality at optimality
under the energy detection scheme and single-user scenario. This is quite intuitive since lower detection probability
implies smaller sensing time, which leads to higher throughput. This is still the case for our considered multi-user
scenario as can be verified by the conditional throughput formula (\ref{con_T}). Therefore, we can set
$\mathcal{P}_d^j\left( {\vec \varepsilon ^j}, {\vec \tau^j} , a_j  \right) = \mathcal{\widehat{P}}_d ^j$ to solve
the optimization problem (\ref{probp1})-(\ref{probp3}).

However, $\mathcal{P}_d^j \left( {\vec \varepsilon ^j}, {\vec \tau^j} , a_j  \right)$ is a function of $\mathcal{P}_d^{ij}$
for all SUs $i \in \mathcal{S}_j^U$ since we employ the SDCSS scheme in this paper. Therefore, to simplify the optimization 
we set $\mathcal{P}_d^{ij}=\mathcal{{P}}_d^{j*}$ for all SUs $i \in \mathcal{S}_j^U$ (i.e., all SUs are required to achieve the same detection probability
for each assigned channel). Then, we can calculate $\mathcal{{P}}_d^{j*}$ by using (\ref{eq1_css_1}) for a given value of $\mathcal{\widehat{P}}_d ^j$. In addition, 
we can determine $\mathcal{P}_f^{ij}$ with the obtained value of $\mathcal{{P}}_d^{j*}$ by using (\ref{eq2}), which is the function of sensing time $\tau^{ij}$.

Even after these steps, the optimization problem (\ref{probp1})-(\ref{probp3}) is still very difficult to solve. In fact, it is
the mixed integer non-linear problem since the optimization variables $a_j$ take integer values while other variables take real values.
Moreover, even the corresponding optimization problem achieved by relaxing $a_j$ to real variables is a difficult and non-convex problem to
solve since the throughput in the objective function (\ref{probp1}) given in (\ref{NTSC_41}) is a complicated and non-linear function of optimization variables.

\begin{algorithm}[h]%\leesize
\caption{\textsc{Optimization of Sensing and Access Parameters }}
\label{mainalg_pg}
%\algsetup{indent=1.5em}
\begin{algorithmic}[1]

\STATE Assume we have the sets of all SU $i$, $\left\{\mathcal{S}_i\right\}$. Initialize $\tau^{ij}$, $j \in \mathcal{S}_i$, the sets of $\left\{a_j\right\}$ for all channel $j$ and $p$.

\STATE For each chosen $p \in \left[0, 1\right]$, find ${\bar \tau}^{ij}$ and $\left\{{\bar a}_j\right\}$ as follows:

\FOR {each possible set $\left\{a_j\right\}$}

\REPEAT

\FOR  {$i = 1$ \text{to} $N$}

\STATE Fix all $\tau^{i_1j}$, $i_1 \neq i$.

\STATE Find the optimal ${\bar \tau}^{ij}$ as ${\bar \tau}^{ij} = \mathop {\argmax} \limits_{0 < \tau^{ij} \leq T} \mathcal{NT}_p\left( \left\{\tau^{ij}\right\}, \left\{a_j\right\}, p\right)$.

\ENDFOR

\UNTIL {convergence}

\ENDFOR

\STATE The best $\left(\left\{{\bar \tau}^{ij}\right\}, \left\{{\bar a}_j\right\}\right)$ is determined for each value of $p$ as $\left(\left\{{\bar \tau}^{ij}\right\}, \left\{{\bar a}_j\right\}\right) = \mathop {\argmax} \limits_{\left\{{ a}_j\right\}, \left\{{\bar \tau}^{ij} \right\}} \mathcal{NT} \left({\bar \tau}^{ij}, \left\{a_j\right\}, p\right)$.

\STATE The final solution $\left( \left\{{\bar \tau}^{ij}\right\}, \left\{{\bar a}_j\right\}, {\bar p}  \right)$ is determined as $\left( \left\{{\bar \tau}^{ij}\right\}, \left\{{\bar a}_j\right\}, {\bar p}  \right) = \mathop {\argmax} \limits_{ \left\{{\bar \tau}^{ij}\right\}, \left\{{\bar a}_j\right\}, p } \mathcal{NT} \left(\left\{{\bar \tau}^{ij}\right\}, \left\{{\bar a}_j\right\}, p\right)$.

\end{algorithmic}
\end{algorithm}

Given this observation, we have devised Alg. \ref{mainalg_pg} to determine the solution for this optimization problem based on the coordinate-descent searching techniques. 
The idea is that at one time we fix all variables while searching for the optimal value of the single variable. This operation is performed sequentially for 
all variables until convergence is achieved. Since the normalized throughput given in (\ref{NTSC_41}) is quite insensitive with respect to $p$, we attempt to determine the 
optimized values for $\left(\left\{{\bar \tau}^{ij}\right\}, \left\{{\bar a}_j\right\} \right)$ first for different values of $p$ (steps 3--11 in Alg.~\ref{mainalg_pg}) before searching
the optimized value of $p$ in the outer loop (step 12 in Alg.~\ref{mainalg_pg}). This algorithm converges to the fixed point solution since we improve
 the objective value over iterations (steps 4--9). 
This optimization problem is non-convex in general. However, we can obtain its optimal solution easily by using the bisection search
technique since the throughput function is quite smooth \cite{Fan10}.
For some specific cases such as in homogeneous systems \cite{Le11, Peh09, Wei09}, the underlying optimization problem is convex, which can be solved efficiently by
using standard convex optimization algorithms.

\subsection{Optimization of Channel Sensing Sets}
\label{GACAPp}

For the CRNs considered in the current work, the network throughput strongly depends on the availability of different channels, the spectrum sensing time, and the
sensing quality. Specifically, long sensing time $\tau$ reduces the communications time on the available channels in each cycle of length $T$, which, therefore,
decreases the network throughput. In addition, poor spectrum sensing performance can also degrade the network throughput since SUs can either overlook
available channels (due to false alarm) or access busy channels (due to missed detection). Thus, the total throughput of SUs can be enhanced by optimizing the
access parameter $p$ and sensing design, namely optimizing the assignments of channels to SUs (i.e., optimizing the sensing sets for SUs)
and the corresponding sensing times.

Recall that we have assumed the channel sensing sets for SUs are fixed to optimize the sensing and access parameters in the previous section.
In this section, we attempt to determine an efficient channel assignment solution (i.e., channel sensing sets) by solving the following problem
\beqn
\label{eq11a}
\max \limits_{ \left\{\mathcal{S}_i\right\}, \left\{a_j\right\}} \mathcal{NT} \left( \left\{{\bar \tau}^{ij}\right\}, \left\{a_j\right\}, {\bar p},  
\left\{\mathcal{S}_i\right\} \right).
\eeqn
Note that the optimal values of $a_j$ can only be determined if we have fixed the channel sensing set $\mathcal{S}_j^U$ for each channel $j$. This is because
we aim to optimize the $a_j$-out-of-$b_j$ aggregation rule of the SDCSS scheme for each channel $j$ where $b_j = |\mathcal{S}_j^U|$. Since $a_j$ takes integer values 
and optimization of channel sensing sets $\mathcal{S}_j^U$ also involves integer variables where we have to determine the set of SUs $\mathcal{S}_j^U$
assigned to sense each channel $j$. Therefore, the optimization problem (\ref{eq11a}) is the non-linear integer program, which is NP-hard \cite{Lee12}.
In the following, we present both brute-force search algorithm and low-complexity greedy algorithm to solve this problem.

\subsubsection{Brute-force Search Algorithm}
\label{BFSA}

Due to the non-linear and combinatorial structure of the formulated channel assignment problem, it would be impossible to explicitly determine the optimal closed form solution
for problem (\ref{eq11a}). However, we can employ the brute-force search (i.e., the exhaustive search) to determine the best channel assignment. 
Specifically, we can enumerate all possible channel assignment solutions. Then, for each channel assignment solution (i.e., sets $\mathcal{S}_j^U$ for all
channels $j$), we employ Alg.~\ref{mainalg_pg} to determine the best spectrum sensing and accessing parameters  $\left\{\tau^{ij}\right\}, \left\{a_j\right\}, p$ and
calculate the corresponding total throughput by using the throughput analytical model in \ref{TputPCSMA}. The channel assignment achieving the
maximum throughput together with its best spectrum sensing and accessing parameters provides the best solution for the optimization problem (\ref{eq11a}).

\subsubsection{Low-Complexity Greedy Algorithm}
\label{PGA}

We propose another low-complexity and greedy algorithm to find the solution for this problem, which is described in  Alg.~\ref{ChanAAp}. 
In this algorithm, we perform the initial channel assignment in step 1, which works as follows. We first temporarily assign all channels for each SU. 
Then, we run Alg. \ref{mainalg_pg} to find the optimal sensing times for this temporary assignment, i.e., to determine $\left\{{\bar \tau}^{ij}\right\}$, which
is used to assign one SU to each channel so that the total sensing time is minimized. In particular, the initial channel assignments are set according to the solution
of the optimization problem (\ref{probp10})-(\ref{probp11}) presented in the following. 
\beqn
 \mathop {\min} \limits_{ \left\{x_{ij}\right\} } \quad \sum_{i,j} \overline{\tau}^{ij} x_{ij} \hspace{1cm} \label{probp10}\\ 
 \mbox{s.t.} \,\,\,\, \sum_{i} x_{ij} = 1, \: j \in \left[1, M\right]. \hspace{0cm} \label{probp11}
\eeqn
where $x_{ij}$ are binary variables representing the channel assignments where
$x_{ij}=1$ if channel $j$ is allocated for SU $i$ (i.e., $j \in \mathcal{S}_i$) and $x_{ij}=0$, otherwise.
We employ the well-known Hungarian algorithm \cite{Kuhn55} to solve this problem. 
Then, we perform further channel assignments in steps 2-18 of Alg.~\ref{ChanAAp}.
Specifically, to determine one channel assignment in each iteration, we temporarily
assign one channel to the sensing set $\mathcal{S}_i$ of each SU $i$ and calculate
the increase of throughput for such channel assignment $\Delta T_{ij}$ with the optimized channel and access parameters
obtained by using Alg.~\ref{mainalg_pg} (step 6). We then
search for the best channel assignment $\left({\bar i}, {\bar j}\right) = \mathop {\argmax} \limits_{i, j \in \mathcal{S} \backslash \mathcal{S}_i} 
\Delta T_{ij} $ and actually perform the corresponding channel assignment if $\Delta T_{\bar i \bar j} >\delta$ (steps 7--10).

In Alg.~\ref{ChanAAp}, $\delta>0$ is a small number which is used in the stopping condition for this algorithm (step 11). 
In particular, if the increase of the normalized throughput due to the new channel assignment is negligible in any iteration
 (i.e., the increase of throughput is less than $\delta$) then the algorithm terminates. Therefore, we can choose
$\delta$ to efficiently balance the achievable throughput performance with the algorithm running time.  
In the numerical studies, we will choose $\delta $ equal to $10^{-3} \times {\mathcal NT}_c $.

The convergence of Alg.~\ref{ChanAAp} can be explained as follows. Over the course of this algorithm, we attempt
to increase the throughput by performing additional channel assignments. It can be observed that
we can increase the throughput by allowing i) SUs to achieve better sensing performance or ii) SUs to reduce their sensing times.
However, these two goals could not be achieved concurrently due to the following reason. 
If SUs wish to improve the sensing performance via cooperative spectrum sensing, we should assign more channels to each of them.
However, SUs would spend longer time sensing the assigned channels with the larger sensing sets, which would ultimately  decrease the throughput.
Therefore, there would exist a point when we cannot improve the throughput by performing further channel assignments, which implies that Alg.~\ref{ChanAAp} must converge.

There is a key difference in the current work and \cite{Le12} regarding the sensing sets of SUs.
Specifically,  the sets of assigned channels are used for spectrum sensing and access in \cite{Le12}.
However, the sets of assigned channels are used for spectrum sensing only in the current work. In addition,
the sets of available channels for possible access at SUs are determined based on the reporting results, which
may suffer from communications errors. 
We will investigate the impact of reporting errors on the throughput performance in Section ~\ref{Exten}.

\begin{algorithm}[!t]%[h]%\leesize
\caption{\textsc{Greedy Algorithm}}
\label{ChanAAp}
%\algsetup{indent=1.5em}
\begin{algorithmic}[1]

\STATE  Initial channel assignment is obtained as follows:

\begin{itemize}

\item Temporarily perform following channel assignments $\widetilde{\mathcal{S}}_i = \mathcal{S}$, $i \in \left[1, N\right]$. Then, run Alg. \ref{mainalg_pg} to obtain
 optimal sensing and access parameters $\left( \left\{{\bar \tau}^{ij}\right\},  \left\{{\bar a}_j\right\}, {\bar p} \right)$. 

\item Employ Hungarian algorithm \cite{Kuhn55} to allocate each channel to exactly one SU to minimize the total cost
where the cost of assigning channel $j$ to SU $i$ is ${\bar \tau}^{ij}$ (i.e., to solve the optimization problem (\ref{probp10})-(\ref{probp11})).

\item The result of this Hungarian algorithm is used to build the initial channel assignment sets $\left\{\mathcal{S}_i\right\}$
for different SU $i$. 
\end{itemize}

\STATE Set $\text{continue} = 1$. %, $k=1$.

\WHILE {\text{continue} = 1}

\STATE Optimize sensing and access parameters for current channel assignment solution $\left\{\mathcal{S}_i \right\}$ by using Alg. \ref{mainalg_pg}.

\STATE Calculate the normalized throughput $\mathcal{NT}_{\sf c} = \mathcal{NT} \left( \left\{{\bar \tau}^{ij}\right\}, \left\{{\bar a}_j\right\}, {\bar p}, 
\left\{\mathcal{S}_i \right\} \right)$ for the optimized sensing and access parameters.

\STATE Each SU $i$ calculates the increase of throughput if it is assigned one further potential channel $j$ as $\Delta T_{ij} = \mathcal{NT} \left( \left\{{\bar \tau}^{ij}\right\}, \left\{{\bar a}_j\right\}, {\bar p}, \left\{\widetilde{\mathcal{S}}_i \right\}\right) - \mathcal{NT}_{\sf c}$ where $\widetilde{\mathcal{S}}_i = \mathcal{S}_i \cup j$,
 $\widetilde{\mathcal{S}}_l = \mathcal{S}_l, \: l \neq i$, and 
$\left\{{\bar \tau}^{ij}\right\}, \left\{{\bar a}_j\right\}, {\bar p}$ are determined
by using Alg. \ref{mainalg_pg} for the temporary assignment sets  $\left\{\widetilde{\mathcal{S}}_i \right\}$.

\STATE Find the ``best'' assignment $\left({\bar i}, {\bar j}\right)$ as $\left({\bar i}, {\bar j}\right) = \mathop {\argmax} \limits_{i, j \in \mathcal{S} \backslash \mathcal{S}_i} \Delta T_{ij} $.

\IF {$\Delta T_{\bar i \bar j} >\delta$}

\STATE Assign channel $\bar j$ to SU $\bar i$: ${\mathcal{S}}_{\overline{i}} = \mathcal{S}_{\overline{i}} \cup \overline{j}$.

\ELSE

\STATE Set $\text{continue} =0$.

\ENDIF

\ENDWHILE

\IF {$\text{continue} =1$}

\STATE Return to step 2.

\ELSE 

\STATE Terminate the algorithm.

\ENDIF

\end{algorithmic}
\end{algorithm}

\subsection{Complexity Analysis}
\label{ComAnap}

In this section, we analyze the complexity of the proposed brute-force search and low-complexity greedy algorithms.

\subsubsection{Brute-force Search Algorithm}
\label{ComAna1p}

To determine the complexity of the brute-force search algorithm, we need to calculate the 
number of possible channel assignments. Since each channel can be either allocated or not allocated
to any SU, the number of channel assignments is $2^{MN}$. Therefore, the complexity of the brute-force search algorithm is $\mathcal{O}\left(2^{MN}\right)$. 
Note that to obtain the best channel assignment solution, we must run Alg. \ref{mainalg_pg} to find the best sensing and access parameters for each potential
channel assignment, calculate the throughput achieved by such optimized configuration, and compare all the throughput values to determine the best solution.

\subsubsection{Low-complexity Greedy Algorithm}
\label{ComAna2p}

In step 1, we run Hungarian algorithm to perform the first channel assignment for each SU $i$. 
The complexity of this operation can be upper-bounded by $\mathcal{O}\left(M^2N\right)$ (see \cite{Kuhn55} for more details). 
In each iteration in the assignment loop (i.e., steps 2-18), each SU $i$ needs to calculate the increases of throughput for different potential channel assignments. 
Then, we select the assignment resulting in maximum increase of throughput.
Hence, the complexity involved in these tasks is upper-bounded by $MN$ since there are at most $M$ channels to assign for each of $N$ SUs. 
Also, the number of assignments to perform is upper bounded by $MN$ (i.e., iterations of the main loop). 
Therefore, the complexity of the assignment loop is upper-bounded by $M^2N^2$. 
Therefore, the total worst-case complexity of Alg. \ref{ChanAAp} is $\mathcal{O}\left(M^2N+M^2N^2\right) = \mathcal{O}\left(M^2N^2\right)$, which is much lower than that of the brute-force search algorithm.
As a result, Table ~\ref{table1} in Section ~\ref{Results} demonstrates that our proposed greedy algorithms achieve the throughput performance very close to that achieved by the brute-force search algorithms albeit they require much lower computational complexity.

\subsection{Practical Implementation Issues}
\label{Prac_Imp}

In our design, the spectrum sensing and access operation is distributed, however, channel assignment is performed in centralized manner.
In fact, one SU is pre-assigned as a cluster head, which conducts channel assignment for SUs (i.e., determine channel sensing
sets for SUs).
For fairness, we can assign the SU as the cluster head in the round-robin manner. 
To perform channel assignment, the cluster head is responsible for estimating $\mathcal{P}_j\left(\mathcal{H}_0\right)$.
Upon determining the channel sensing sets for all SUs, the cluster head will forward the results to the SUs.
Then based on these pre-determined sensing sets, SUs will perform spectrum sensing and run the underlying MAC protocol to access the channel distributively in each cycle.
It is worth to emphasize that the sensing sets for SUs are only determined once the probabilities $\mathcal{P}_j\left(\mathcal{H}_0\right)$ change, which would be quite infrequent in practice (e.g., in the time scale of hours or even days). 
Therefore, the estimation cost for $\mathcal{P}_j\left(\mathcal{H}_0\right)$ and all involved communication overhead due to sensing set optimization operations would be acceptable.

\vspace{10pt}
\section{Consideration of Reporting Errors}
\label{Exten}

In this section, we consider the impact of reporting errors on the performance of the proposed joint SDCSS and access design.
Note that each SU relies on the channel sensing results received from other SUs in $\mathcal{S}_j^U$ to determine the sensing
outcome for each channel $j$. If there are reporting errors then different SUs may receive
different channel sensing results, which lead to different final channel sensing decisions. The throughput analysis, therefore,
must account for all possible error patterns that can occur in reporting channel sensing results.  
We will present the cooperative sensing model and throughput analysis considering reporting errors in the following.

\subsection{Cooperative Sensing with Reporting Errors}
\label{CS_w_RE}

In the proposed SDCSS scheme, each SU $i_1 $ collects sensing results for each channel $j $ from all SUs $i_2 \in \mathcal{S}_j^U $ who are assigned to sense channel $j$.
In this section, we consider the case where there can be errors in reporting the channel sensing results among SUs. We assume that the channel sensing
result for each channel transmitted by one SU to other SUs is represented by a single bit whose 1/0 values indicates that the underlying channel is available and busy, respectively.
In general, the error probability of the reporting message between SUs $i_1$ and $i_2$ depends on the employed modulation scheme and the signal to noise ratio (SNR) of the
communication channel between the two SUs. We denote the bit error probability of transmitting the reporting bit from SU  $i_2$ to SU $i_1$ as $\mathcal{P}_e^{i_1i_2} $. 
In addition, we assume that the error processes of different reporting bits for different SUs are independent.
Then, the probability of detection and probability of false alarm experienced by SU $i_1$ on channel $j$ with the sensing result received from SU $i_2$ can be 
written as 
\beqn
\mathcal{P}_{u,e}^{i_1i_2j} \! = \! \left\{\!\!\! \begin{array}{*{20}{c}}
   \mathcal{P}_u^{i_2j} \left(1 \! - \! \mathcal{P}_e^{i_1i_2}\right) \! + \! \left(1 \! - \! \mathcal{P}_u^{i_2j}\right) \mathcal{P}_e^{i_1i_2} & {\mbox{if } i_1 \! \neq \! i_2}  \\
   \mathcal{P}_u^{i_2j}  & {\mbox{if } i_1 \! = \! i_2}  \\
\end{array} \!\!\! \right.  \label{eq1_dcss_1}
\eeqn
where $u \equiv d$ and $u \equiv f$ represents probabilities of detection and false alarm, respectively. 
Note that we have $\mathcal{P}_e^{i_1i_2} = 0$ if $i_1=i_2=i$ since there is no sensing result exchange involved in this case.
As SU $i$ employs the $a_j$-out-of-$b_j$ aggregation rule for channel $j$, the probabilities of detection and false alarm for SU $i$ on channel $j$ 
can be calculated as
\beqn
\mathcal{\tilde{P}}_u^{ij}\left( {\vec \varepsilon ^j}, {\vec \tau^j} , a_j  \right) = \sum_{l=a_j}^{b_j} \sum_{k=1}^{C_{b_j}^l} \prod_{i_1 \in \Phi^l_k} \mathcal{P}_{u,e}^{ii_1j} \prod_{i_2 \in \mathcal{S}_j^{U} \backslash \Phi^l_k} \mathcal{\bar P}_{u,e}^{ii_2j}. \label{Pu_rep_1}
\eeqn
Again, $u \equiv d$ and $u \equiv f$ represent the corresponding probabilities of detection or  false alarm, respectively. 
Recall that $\mathcal{S}_j^U $ represents the set of SUs who are assigned to sense channel $j$; thus, we have $b_j = |\mathcal{S}_j^U|$
and $1 \leq a_j \leq b_j = \left| \mathcal{S}_j^U \right|$.    
For brevity, $\mathcal{\tilde P}_u^{ij}\left({\vec \varepsilon ^j}, {\vec \tau^j}, a_j  \right) $ is written as $\mathcal{\tilde P}_u^{ij}$ in the following.

\subsection{Throughput Analysis Considering Reporting Errors}
\label{TputanaCRE}

\begin{figure}[!t]%[!t]
\centering
\includegraphics[width=85mm]{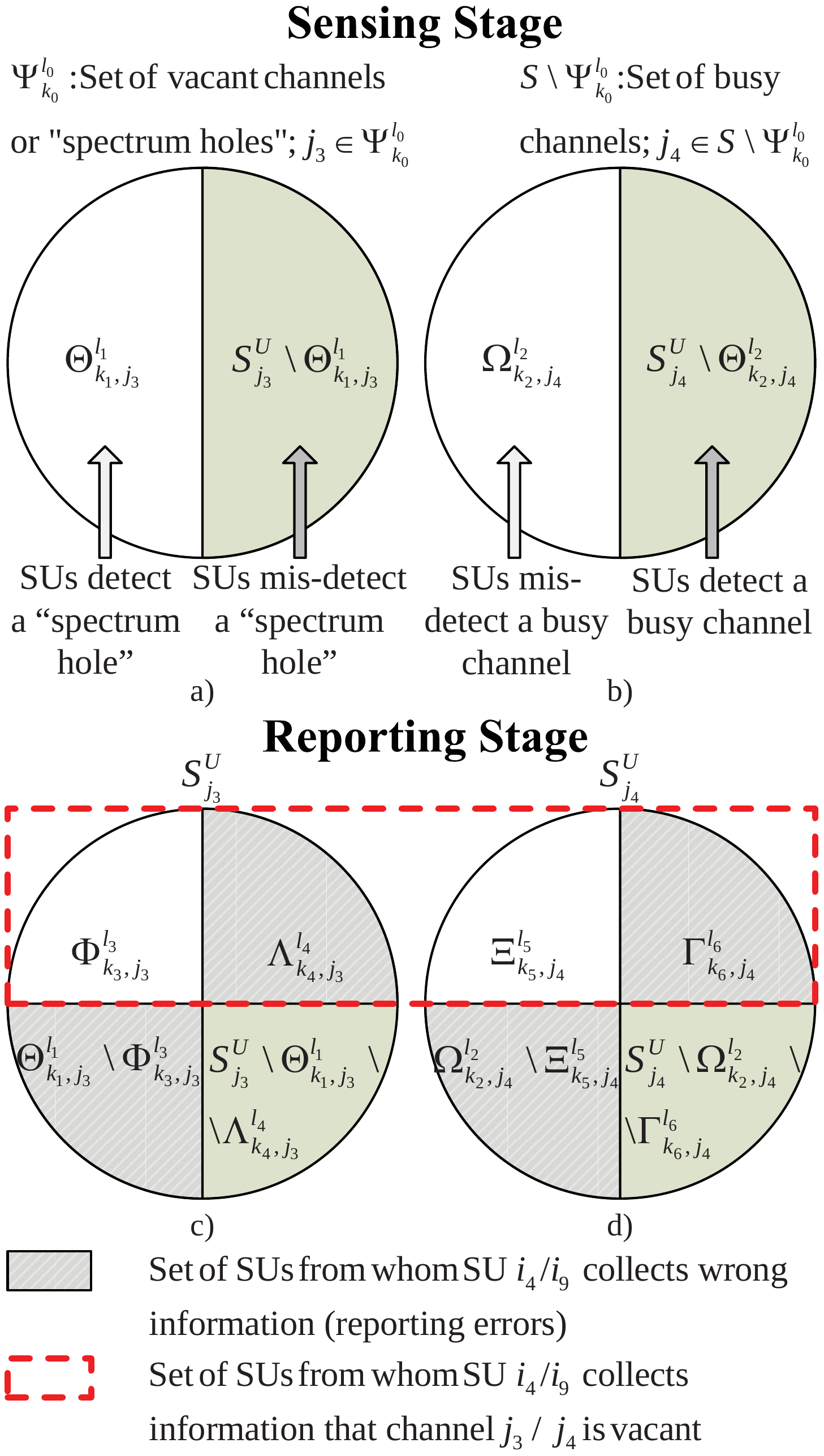}
\caption{Illustration of different sets in one combined scenario.}
\label{Fig0}
\end{figure}

In order to analyze the saturation throughput for the case there are reporting errors, we have to
consider all possible scenarios due to the idle/busy status of all channels, sensing outcomes given by different SUs, and error/success 
events in the sensing result exchange processes. For one such combined scenario we have to derive the total conditional
throughput due to all available channels. Illustration of different involved sets for one combined scenario of following analysis is presented in Fig.~\ref{Fig0}.
In particular, the normalized throughput considering reporting errors can be expressed as follows:
\beqn
\mathcal{NT} = \sum_{k_0=1}^M \sum_{l_0=1}^{C_M^{k_0}} \prod_{j_1 \in \Psi_{k_0}^{l_0}} \mathcal{P}_{j_1} \left(\mathcal{H}_0\right) \prod_{j_2 \in \mathcal{S} \backslash \Psi_{k_0}^{l_0}} \mathcal{P}_{j_2} \left(\mathcal{H}_1\right) \times \label{NTSC_1}\\
\prod_{j_3 \in \Psi_{k_0}^{l_0}} \sum_{k_1=0}^{ |\mathcal{S}_{j_3}^U| } \sum_{l_1=1}^{C_{|\mathcal{S}_{j_3}^U|}^{k_1}} \prod_{i_0 \in \Theta_{k_1,j_3}^{l_1}} 
\mathcal{\bar P}_f^{i_0,j_3} 
\prod_{i_1 \in  \mathcal{S}_{j_3}^U \backslash \Theta_{k_1,j_3}^{l_1}} \mathcal{P}_f^{i_1,j_3} \times \label{NTSC_2} \\
\prod_{j_4 \in \mathcal{S} \backslash \Psi_{k_0}^{l_0}} \sum_{k_2 = 0}^{|\mathcal{S}_{j_4}^U|} \sum_{l_2=1}^{C_{|\mathcal{S}_{j_4}^U|}^{k_2}} \prod_{i_2 \in \Omega_{k_2,j_4}^{l_2}} \mathcal{\bar P}_d^{i_2,j_4} \prod_{i_3 \in \mathcal{S}_{j_4}^U \backslash \Omega_{k_2,j_4}^{l_2}} \mathcal{P}_d^{i_3,j_4} \times  \label{NTSC_3} \\
\prod_{i_4 \in  \mathcal{S}^U }  \sum_{k_3=0}^{k_1 } \sum_{l_3=1}^{C_{k_1 }^{k_3}} \prod_{i_5 \in \Phi_{k_3,j_3}^{l_3}}  \mathcal{\bar P}_e^{i_4,i_5} \prod_{i_6 \in \Theta_{k_1,j_3}^{l_1} \backslash \Phi_{k_3,j_3}^{l_3}}  \mathcal{P}_e^{i_4,i_6} \times \label{NTSC_40} \\
\sum_{k_4=0 }^{|\mathcal{S}_{j_3}^U| - k_1 } \sum_{l_4=1}^{C_{|\mathcal{S}_{j_3}^U| - k_1 }^{k_4 }} \prod_{i_7 \in \Lambda_{k_4,j_3}^{l_4 }}  \mathcal{P}_e^{i_4,i_7} \!\!\!\!\!\prod_{i_8 \in \mathcal{S}_{j_3}^U \backslash \Theta_{k_1,j_3}^{l_1} \backslash \Lambda_{k_4,j_3}^{l_4}} \!\!\!\!\!  \mathcal{\bar P}_e^{i_4,i_8} \times \label{NTSC_4} \\
\prod_{i_9 \in  \mathcal{S}^U }  \sum_{k_5=0}^{k_2 } \sum_{l_5=1}^{C_{k_2 }^{k_5}} \prod_{i_{10} \in \Xi_{k_5,j_4}^{l_5}}  \mathcal{\bar P}_e^{i_9,i_{10}} \!\!\!\! \prod_{i_{11} \in \Omega_{k_2,j_4}^{l_2} \backslash \Xi_{k_5,j_4}^{l_5}}  \!\!\!\! \mathcal{P}_e^{i_9,i_{11}} \times  \label{NTSC_50} \\
\sum_{k_6=0 }^{|\mathcal{S}_{j_4}^U| - k_2 } \sum_{l_6=1}^{C_{|\mathcal{S}_{j_4}^U| - k_2 }^{k_6 }} \prod_{i_{12} \in \Gamma_{k_6,j_4}^{l_6 }} \!\!\!\! \mathcal{P}_e^{i_9,i_{12}} \!\!\!\!\!\!\!\! \prod_{i_{13} \in \mathcal{S}_{j_4}^U \backslash \Omega_{k_2,j_4 }^{l_2 } \backslash \Gamma_{k_6,j_4}^{l_6 }} \!\!\!\!\!\!\!\! \mathcal{\bar P}_e^{i_9,i_{13}} \times \label{NTSC_5} \\
\mathcal{T}_p^{\sf re} \left(\tau,\left\{a_j\right\},p  \right), \label{NTSC_6}
\eeqn
%\beqn
%\eeqn
where $\mathcal{T}_p^{\sf re} \left(\tau,\left\{a_j\right\},p  \right)$ denotes the conditional throughput for one
combined scenario discussed above. 
In (\ref{NTSC_1}), we generate all possible sets where $k_0 $ channels are available for secondary access (i.e., they are not used by PUs)
while the remaining channels are busy. There are $C_{M}^{k_0}$ such sets and $\Psi_{k_0}^{l_0} $ represents one particular set of available channels. 
The first product term in (\ref{NTSC_1}) denotes the probability that all channels in $\Psi_{k_0}^{l_0} $ are available while the second product term describes
 the probability that the remaining channels are busy. %Note that $\bar{\Psi}_{k_0}^{l_0} = \mathcal{S} \backslash \Psi_{k_0}^{l_0}$. 

Then, for one particular channel $j_3 \in \Psi_{k_0}^{l_0}$, we generate all possible sets with $k_1 $ SUs in $\mathcal{S}_{j_3}^U $ ($\mathcal{S}_{j_3}^U $ is the set of SUs 
who are assigned to sense channel $j_3 $) whose sensing results indicate that channel $j_3 $ is available in (\ref{NTSC_2}).
There are $C_{\left|\mathcal{S}_{j_3}^U\right|}^{k_1 } $ sets and $\Theta_{k_1,j_3}^{l_1} $ denotes one such typical set.
Again, the first product term in (\ref{NTSC_2}) is the probability that the sensing outcomes of all SUs in $\Theta_{k_1,j_3}^{l_1} $ indicate that channel $j_3 $ is available; and 
the second term is the probability that the sensing outcomes of all SUs in the remaining set $\mathcal{S}_{j_3}^U \backslash \Theta_{k_1,j_3}^{l_1} $ indicate 
that channel $j_3 $ is not available.

In (\ref{NTSC_3}), for one specific channel $j_4 \in \mathcal{S} \backslash \Psi_{k_0}^{l_0}$, we generate all possible sets with $k_2 $ SUs in 
$\mathcal{S}_{j_4}^U $ whose sensing outcomes indicate that channel $j_4 $ is available due to missed detection.
There are $C_{\left|\mathcal{S}_{j_4}^U\right|}^{k_2} $ such sets and $\Omega_{k_2,j_4}^{l_2} $ is a typical one.
Similarly, the first product term in (\ref{NTSC_3}) is the probability that the sensing outcomes of all SUs in $\Omega_{k_2,j_4}^{l_2} $ indicate 
that channel $j_4 $ is available; and the second term is the probability that the sensing outcomes of all SUs in the remaining set $\mathcal{S}_{j_4}^U 
\backslash \Omega_{k_2,j_4}^{l_2} $ indicate that channel $j_4 $ is not available.

Recall that for any specific channel $j$, each SU in $\mathcal{S}^U $ (the set of all SUs) receives sensing results from 
a group of SUs who are assigned to sense the channel $j$. In (\ref{NTSC_40}), we consider all possible error events due to message exchanges from 
SUs in $\Theta_{k_1,j_3}^{l_1} $.
The first group denoted as $\Phi_{k_3,j_3}^{l_3}$ includes SUs in $\Theta_{k_1,j_3}^{l_1} $ has its sensing results received at SU  $i_4 \in \mathcal{S}^U $ indicating that 
channel $j_3 $ available (no reporting error) while the second group of SUs $  \Theta_{k_1,j_3}^{l_1} \backslash  \Phi_{k_3,j_3}^{l_3}$ has the sensing results 
received at SU  $i_4 \in \mathcal{S}^U $ suggesting that channel $j_3 $ is not available due to reporting errors. For each of these two groups, we generate all possible
 sets of SUs of different sizes and capture the corresponding
probabilities. In particular, we generate all sets with $k_3 $ SUs $i_5 \in \Phi_{k_3,j_3}^{l_3}$ where SU $i_4 $ collects correct sensing information from SUs $i_5 $ 
(i.e., there is no error on the channel between $i_4 $ and $i_5 $). Similar expression is presented for the second group in which we generate all sets of 
$k_4$ SUs $i_6  \in \Theta_{k_1,j_3}^{l_1} \backslash  \Phi_{k_3,j_3}^{l_3} $ where SU $i_4 $ collects wrong sensing information from each SU $i_6 $ (i.e., there is an error on 
the channel between $i_4 $ and $i_6 $). Similarly, we present the possible error events due to exchanges of sensing results from the set of SUs $\mathcal{S}_{j_3}^U \backslash \Theta_{k_1,j_3}^{l_1} $ in (\ref{NTSC_4}).

In (\ref{NTSC_50}) and (\ref{NTSC_5}), we consider all possible error events due to sensing result exchanges
for channel $j_4 \in \mathcal{S} \backslash \Psi_{k_0}^{l_0}$. Here, each SU in $\mathcal{S}^U $ collects sensing result information from two sets of SUs in $\Omega_{k_2,j_4}^{l_2} $
and $\mathcal{S}_{j_4}^U \backslash \Omega_{k_2,j_4}^{l_2} $, respectively. 
The first set includes SUs in $\Omega_{k_2,j_4}^{l_2} $ whose sensing results indicate that channel $j_4 $ available due to missed detection, while the second set includes 
SUs in $\mathcal{S}_{j_4}^U \backslash \Omega_{k_2,j_4}^{l_2} $ whose sensing results indicate that channel $j_4 $ is not available. 
Possible outcomes for the message exchanges due to the first set $\Omega_{k_2,j_4}^{l_2} $ are captured in (\ref{NTSC_50}) where we present the outcomes
for two groups of this first set. For group one, we generate all sets with $k_5 $ SUs $i_{10} \in \Xi_{k_5,j_4}^{l_5}$ where SU $i_9 $ collects correct sensing information from SUs $i_{10} $ (i.e., there is no error on the channel between $i_9 $ and $i_{10} $). 
For group two, we consider the remaining sets of SUs in $\Omega_{k_2,j_4}^{l_2}  \backslash  \Xi_{k_5,j_4}^{l_5} $ where SU $i_9 $ receives
wrong sensing information from each SU $i_{11} $ (i.e., there is an error on the channel between $i_9 $ and $i_{11} $).
Similar partitioning of the set $\mathcal{S}_{j_4}^U \backslash \Omega_{k_2,j_4}^{l_2} $ into two groups $\Gamma_{k_6,j_4}^{l_6 }$ and $\mathcal{S}_{j_4}^U \backslash \Omega_{k_2,j_4}^{l_2} \backslash \Gamma_{k_6,j_4}^{l_6 }$ with the corresponding message reporting error patterns is captured in (\ref{NTSC_5}).

For each combined scenario whose probability is presented above, each SU $i$ has collected sensing result information for each
channel, which is the sensing results obtained by itself or received from other SUs.
Then, each SU $i$ determines the idle/busy status of each channel $j$ by applying the $a_j$-out-of-$b_j$ rule on the collected sensing information. 
Let $\mathcal{S}^a_i$ be set of channels, whose status is ``available'' as being suggested by the $a_j$-out-of-$b_j$ rule at SU $i$. 
According to our design MAC protocol, SU $i$ will randomly select one channel in the set $\mathcal{S}^a_i$ to perform contention and transmit its data.
In order to obtain the conditional throughput $\mathcal{T}_p^{\sf re} \left(\tau,\left\{a_j\right\},p  \right)$ for one particular
combined scenario, we have to reveal the contention operation on each actually available channel, which is presented in the following.

Let  $\mathcal{S}^a_i =  \mathcal{S}^a_{1,i} \cup \mathcal{S}^a_{2,i} $ where channels in $\mathcal{S}^a_{1,i}$ are actually available and
channels in $\mathcal{S}^a_{2,i}$ are not available but the SDCSS policy suggests the opposite due to sensing and/or reporting errors.
Moreover, let $\mathcal{\hat S}^a_1 = \bigcup_{i \in \mathcal{S}^U} \mathcal{S}^a_{1,i} $ be the set of actually available channels, which are detected by
 all SUs by using the  SDCSS policy. Similarly, we define $\mathcal{\hat S}^a_2 = \bigcup_{i \in \mathcal{S}^U} \mathcal{S}^a_{2,i} $ 
as the set of channels indicated as available by some SUs due to errors. Let $k_e^i = \left|\mathcal{S}^a_i\right|$ be the number of available channels at SU $i $;
then SU $i$ chooses one channel in  $\mathcal{S}^a_i$ to transmit data with probability  $1/k_e^i$.
In addition, let $\mathcal{\hat S}^a = \mathcal{\hat S}^a_1 \cup \mathcal{\hat S}^a_2 $ be set of all ``available'' channels each of which is determined as being available by at 
least one SU and let $k_{\sf{max}} = \left|\mathcal{\hat S}^a\right|$ be the size of this set.

To calculate the throughput for each channel $j$, let $\Psi_j^a $ be the set of SUs whose SDCSS outcomes indicate that channel $j $ is available
and let $\Psi^a = \bigcup_{j \in \hat{\mathcal S}^a} \Psi_j^a $ be the set of SUs whose SDCSS outcomes indicate that at least one channel in the assigned spectrum sensing set is available. 
In addition, let us define $N_j = \left|\Psi_j^a \right|$ and $N_{\sf{max}} = \left|\Psi^a \right|$, which describe the sizes of these sets, respectively.
It is noted that $N_{\sf{max}} \leq N $ due to the following reason. 
In any specific combination that is generated in Eqs. (\ref{NTSC_1})--(\ref{NTSC_5}), there can be some SUs, denoted as $\left\{i\right\}$, whose sensing outcomes 
indicate that all channels in the assigned spectrum sensing sets are not available (i.e., not available for access).
Therefore, we have $\Psi^a = {\mathcal S}^U \backslash \left\{i\right\}$, which implies $N_{\sf{max}} \leq N $ where $N = \left|{\mathcal S}^U \right| $.
Moreover, we assume that channels in $\hat{\mathcal S}^a$ are indexed by $1, 2, \ldots, k_{\sf{max}}$.
Similar to the throughput analysis without reporting errors, we consider all possible sets $\left\{ n_{j} \right\} = \left\{n_1, n_2, \ldots, n_{k_{\sf{max}}} \right\}$
where $n_j$ is the number of SUs choosing channel $j$ for access. Then, we can calculate the conditional throughput as follows:
\beqn
\mathcal{T}_p^{\sf re} \left(\tau,\left\{a_j\right\},p \right)  = \!\!\! \sum_{\left\{n_{j_1}\right\}: \sum _{j_1 \in \mathcal{\hat S}^a } n_{j_1}=N_{\text{max}}} \!\!\!
\mathcal{P}\left(\left\{N_{j_1},n_{j_1}\right\}\right) \times \hspace{0.0cm} \label{T_p_cal_re1} \\
\sum_{j_2 \in \mathcal{\hat S}_1^a} \frac{1}{M} \mathcal{T}_{j_2}^{\sf re} \left(\tau,\left\{a_{j_2}\right\},p \left|n=n_{j_2}\right.\right)   \mathcal{I} \left(n_{j_2}>0\right). \label{T_p_cal_re2} 
\eeqn
Here $\mathcal{P}\left(\left\{N_{j_1},n_{j_1}\right\}\right)$ is the probability that each channel $j_1 $ ($j_1 \in \mathcal{\hat S}^a $) is selected by $n_{j_1} $ SUs
for $j_1=1, 2, \ldots, k_{\text{max}}$. This probability can be calculated as
\beqn
\label{P_Nj1_nj1}
\mathcal{P}\left(\left\{N_{j_1},n_{j_1}\right\}\right) = \left( {\begin{array}{*{20}{c}}
   {\left\{N_{j_1}\right\}}  \\
   {\left\{n_{j_1}\right\}}  \\
\end{array}} \right) \prod_{i \in \Psi^a}\left(\frac{1}{k_e^i}\right),
\eeqn
where  $\left( {\begin{array}{*{20}{c}}
   {\left\{N_{j_1}\right\}}  \\
   {\left\{n_{j_1}\right\}}  \\
\end{array}} \right) $
describes the number of ways to realize the access vector $\left\{ n_{j} \right\}$ for $k_{\sf{max}}$ channels, which can be obtained by using the enumeration technique as follows. 
For a particular way that the specific set of $n_1 $ SUs ${\mathcal S}_{1}^{n_1}$ choose channel one (there are $C^{n_1}_{N_1}$ such ways), we can express the set of remaining SUs 
that can choose channel two as $\Psi^a_{(2)} = \Psi_2^a \backslash ({\mathcal S}_{1}^{n_1} \cap \Psi_2^a) $. 
We then consider all possible ways that $n_2 $ SUs in the set $\Psi^a_{(2)} $ choose channel two and we denote this set of SUs as ${\mathcal S}_{2}^{n_2} $ (there are $C^{n_2}_{\widetilde{N}_2}$ 
such ways where $\widetilde{N}_2 = |\Psi^a_{(2)}|$). Similarly, we can express the set of SUs that can choose channel three as $\Psi^a_{(3)} = \Psi^a_3 \backslash ((\cup_{i=1}^2 {\mathcal S}_{i}^{n_i}) \cap \Psi^a_3) $ and consider all possible ways that $n_3$ SUs in the set $\Psi^a_{(3)}$ can choose channel three, and so on. 
This process is continued until $n_{k_{\sf{max}}}$ SUs choose channel $k_{\sf{max}}$. Therefore, the number of ways to realize the access vector $\left\{ n_{j} \right\}$ can be 
determined by counting all possible cases in the enumeration process.

The product term in (\ref{P_Nj1_nj1}) is due to the fact that each SU $i$ chooses one available with probability $1/k_e^i$. 
The conditional throughput $\mathcal{T}_{j_2}^{\sf re} \left(\tau, \left\{a_{j_2}\right\},p\left| n=n_{j_2}\right. \right) $ is calculated by using the same expression (\ref{con_T})
given in Section \ref{CPCSMA}. In addition, only actually available channel $j_2 \in \mathcal{\hat S}_1^a$ can contribute the total throughput, which explains the throughput 
sum in (\ref{T_p_cal_re2}).

\subsection{Design Optimization with Reporting Errors}

The optimization of channel sensing/access parameters as well as channel sensing sets can be conducted in the same manner with that
in Section \ref{CPCSMA}. However, we have to utilize the new throughput analytical model presented in Section \ref{TputanaCRE} in this case. Specifically, Algs. \ref{mainalg_pg} and \ref{ChanAAp} can 
still be used to determine the optimized sensing/access parameters and channel sensing sets, respectively. Nonetheless,
we need to use the new channel sensing model capturing reporting errors in Section \ref{CS_w_RE} in these algorithms. In particular,
 from the equality constraint on the detection probability, i.e., $\mathcal{P}_d^j\left( {\vec \varepsilon ^j}, {\vec \tau^j} , a_j  \right) = \mathcal{\widehat{P}}_d ^j$,
we have to use (\ref{eq1_dcss_1}) and (\ref{Pu_rep_1}) to determine ${P}_d^{ij}$ (and the corresponding ${P}_f^{ij}$)
assuming that ${P}_d^{ij}$ are all the same for all pairs $\left\{i,j\right\}$ as what we have done in Section \ref{CPCSMA}.

\vspace{10pt}
\section{Numerical Results}
\label{Results}

To obtain numerical results in this section, the key parameters for the proposed MAC protocol are chosen as follows:
cycle time is $T = 100 ms$; the slot size is $v=20{\mu} s$, which is the same as in IEEE 802.11p standard; packet size is $PS = 450$ slots (i.e., $450v$); 
propagation delay $PD = 1 {\mu} s$; $SIFS = 2$ slots; $DIFS = 10 $ slots; $ACK = 20$ slots; $CTS = 20$ slots; $RTS = 20$ slots;
sampling frequency for spectrum sensing is $f_s = 6 MHz$;  and $t_r = 80 {\mu} s$.  %bandwidth of PUs' QPSK signals is $6 MHz$;
The results presented in all figures except Fig. \ref{Fig9} correspond to the case where there is no reporting error.

\begin{table*} % [!t]
\centering
\caption{Throughput vs probability of vacant channel (MxN=4x4)}
\label{table1}
\begin{tabular}{|c|c|c|c|c|c|c|c|c|c|c|c|}
\cline{3-12} 
\multicolumn{2}{c|}{} & \multicolumn{10}{c|}{$\mathcal{P}_j\left(\mathcal{H}_0\right)$}\tabularnewline
\cline{3-12} 
\multicolumn{2}{c|}{} & 0.1 & 0.2 & 0.3 & 0.4 & 0.5 & 0.6 & 0.7 & 0.8 & 0.9 & 1\tabularnewline
\hline 
 & Greedy &0.0816  &  0.1524   & 0.2316   & 0.2982   & 0.3612   & 0.4142  &  0.4662   & 0.5058   & 0.5461  &  0.5742 \tabularnewline
\cline{2-12} 
$\mathcal{NT}$ & Optimal & 0.0817  &  0.1589  &  0.2321  &  0.3007  &  0.3613  &  0.4183  &  0.4681  &  0.5087  &  0.5488  &  0.5796 \tabularnewline
\cline{2-12} 
 & Gap (\%) & 0.12    &    4.09   & 0.22  &  0.83   &     0.03  &  0.98    &     0.40  &  0.57  &  0.49    &   0.93 \tabularnewline
\hline
\end{tabular}
\end{table*}

To investigate the efficacy of our proposed low-complexity channel assignment algorithm (Alg. \ref{ChanAAp}), we compare the throughput performance achieved by the optimal brute-force search and greedy channel assignment algorithm in Table ~\ref{table1}. 
In particular, we show normalized throughput $\mathcal{NT}$ versus probabilities $\mathcal{P}_j\left(\mathcal{H}_0\right)$ for these two algorithms and the relative gap between them. 
Here, the probabilities $\mathcal{P}_j\left(\mathcal{H}_0\right)$ for different channels $j$  are chosen to be the same and we choose $M = 4$ channels and $N = 4$ SUs. 
To describe the SNR of different SUs and channels, we use  $\left\{ i,j \right\}$ to denote a combination of channel $j$ and SU $i$ who senses this channel. 
The SNR setting for different combinations of SUs and channels $\left\{i, j\right\}$ is performed for two groups of SUs as $\gamma_1^{ij} = -15 dB$: channel 1: $\left\{1,1\right\}, \left\{2,1\right\}, \left\{3,1\right\} $; channel 2: $\left\{2,2\right\}, \left\{4,2\right\} $; channel 3: $\left\{1,3\right\},\left\{4,3\right\}$; and channel 4: $\left\{1,4\right\}, \left\{3,4\right\}$. 
The remaining combinations correspond to the SNR value $\gamma_2^{ij} = -20 dB$ for group two. 
The results in this table confirms that the throughput gaps between our greedy algorithm and the brute-force optimal search algorithm are quite small, which are less that 1\% for all except the case two presented in this table. 
These results confirm that our proposed greedy algorithm works well for small systems (i.e., small M and N).
In the following, we investigate the performance of our proposed algorithms for larger systems.

\begin{figure}[!t]
\centering
\includegraphics[width=85mm]{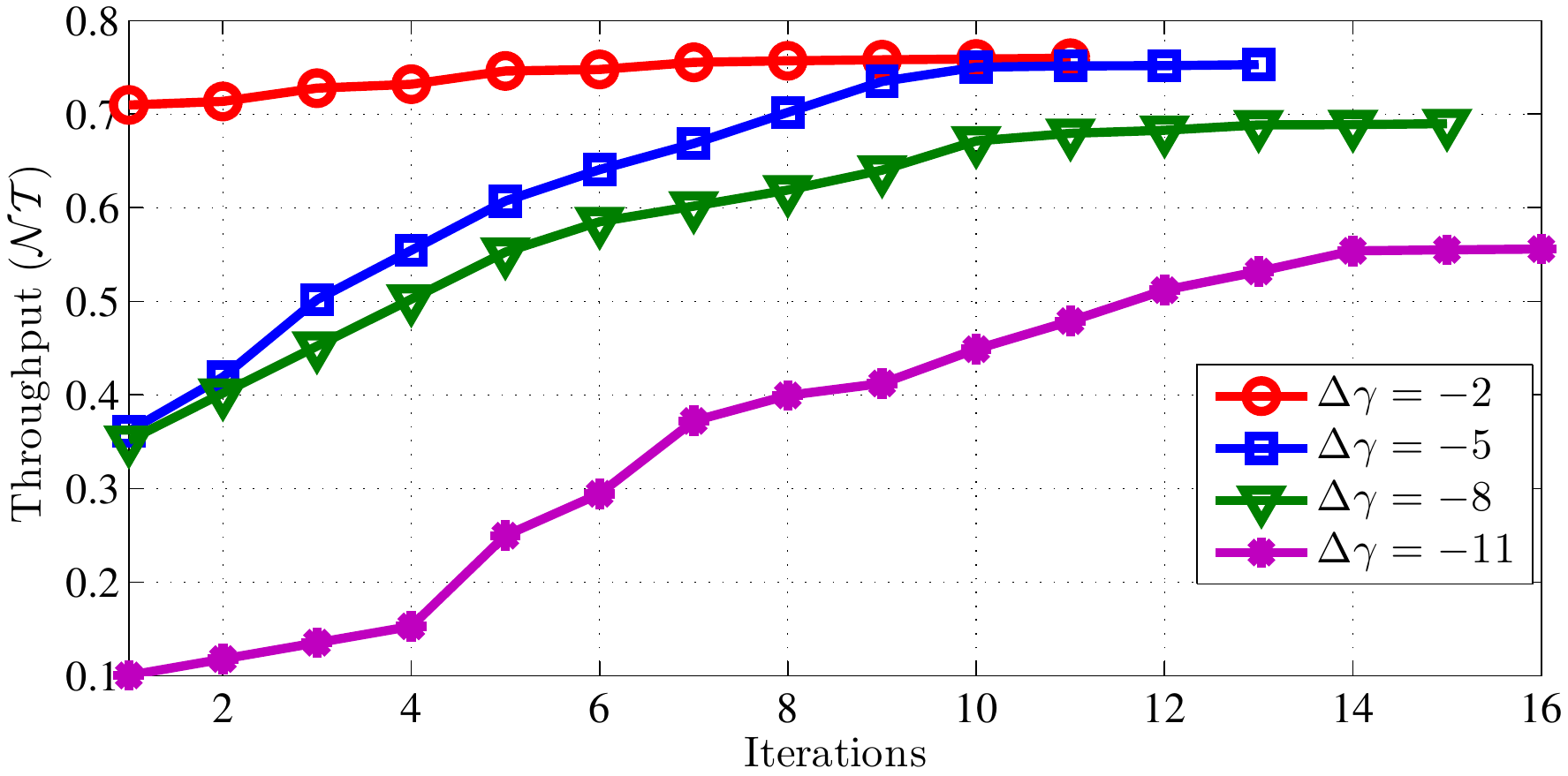}
\caption{Convergence illustration for Alg.~\ref{ChanAAp}.}
\label{Convergence_NT_Iter}
\end{figure}

\begin{figure}[!t]
\centering
\includegraphics[width=85mm]{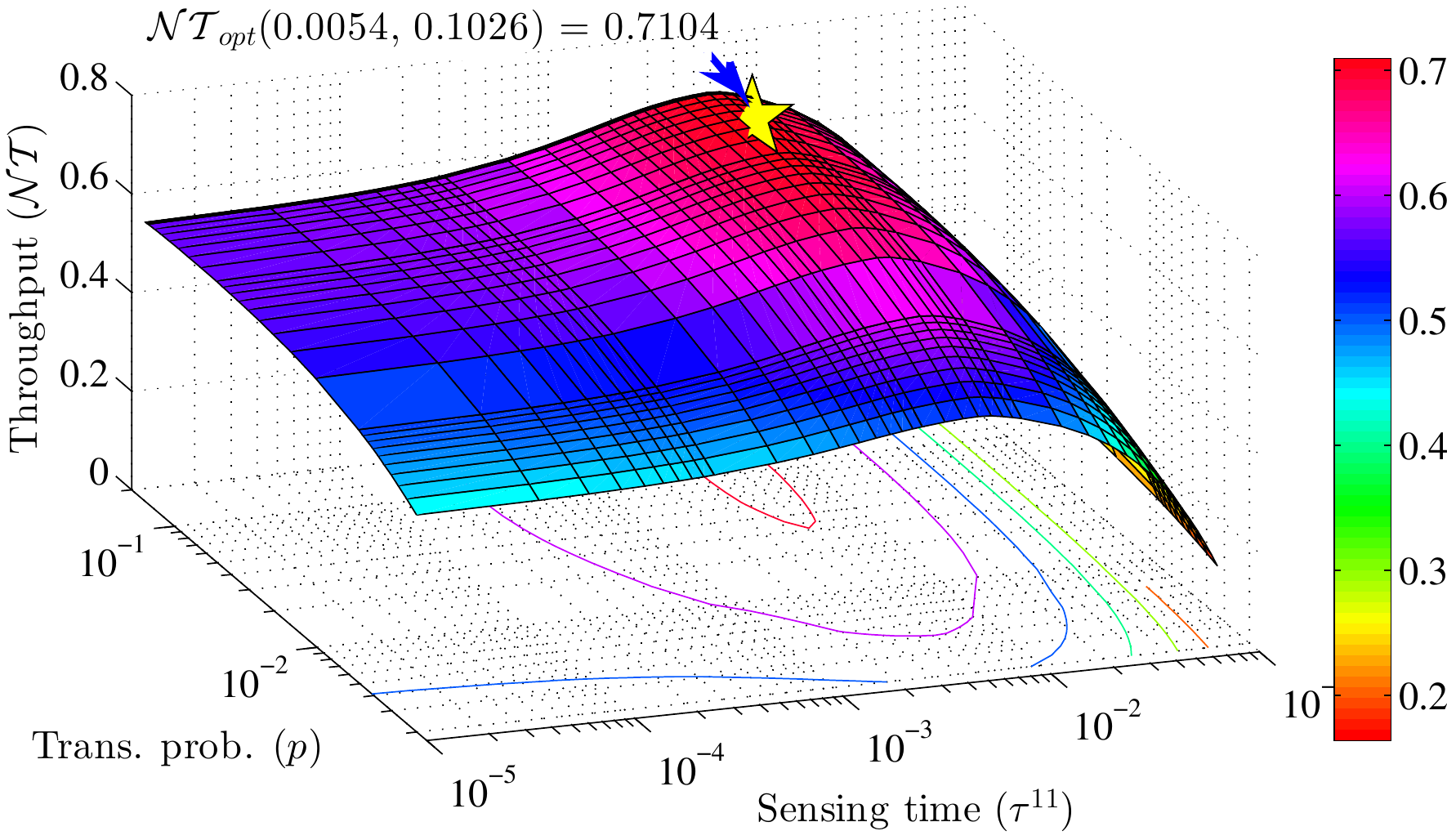}
\caption{Normalized throughput versus transmission probability $p$ and sensing time $\tau ^{11}$ for $\Delta \gamma = -7$, $N=10$ and $M=4$.}
\label{Fig4}
\end{figure}

\begin{figure}[!t]
\centering
\includegraphics[width=85mm]{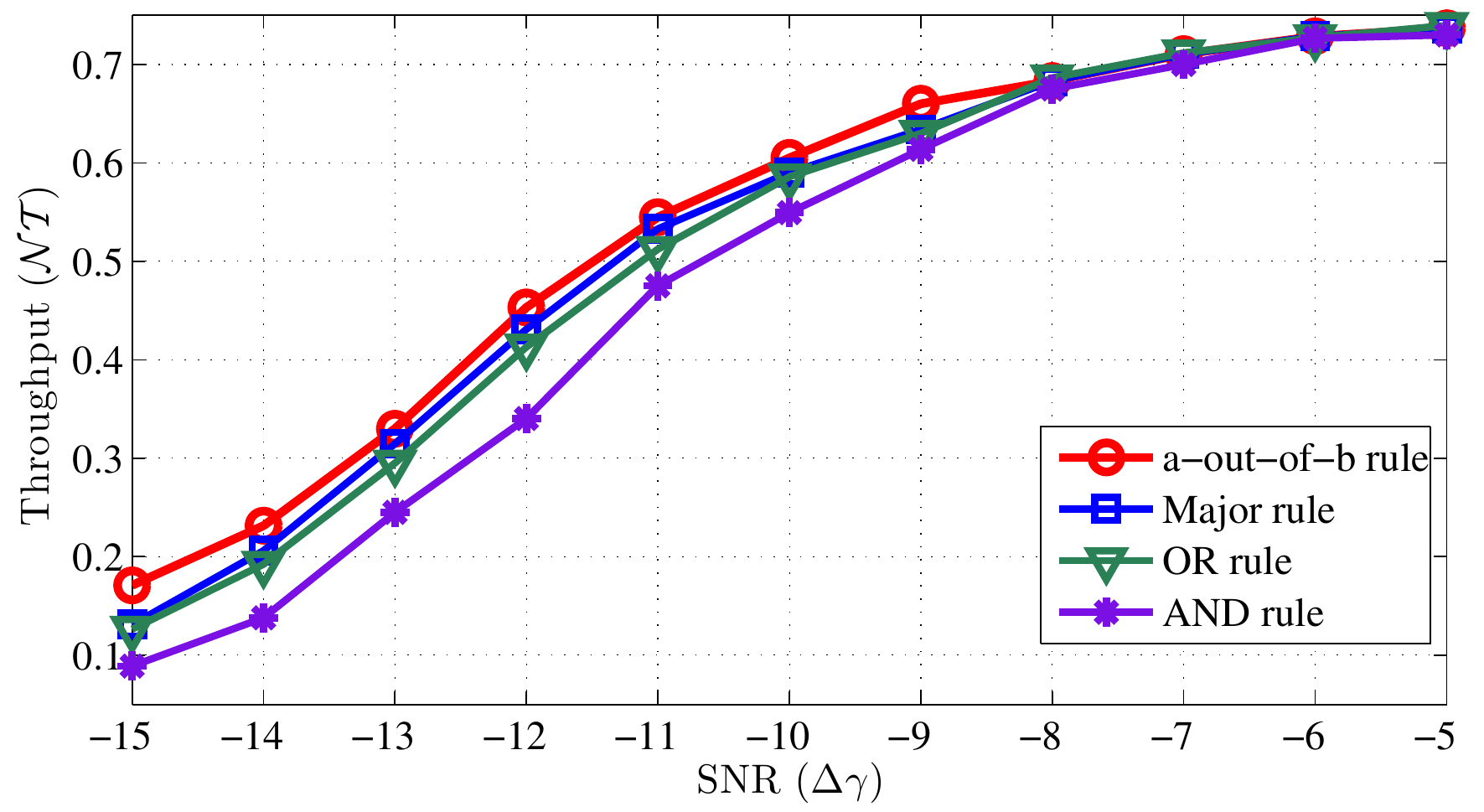}
\caption{Normalized throughput versus SNR shift $\Delta \gamma$ for $N=10$ and $M=4$ under 4 aggregation rules.}
\label{Fig5}
\end{figure}

\begin{figure}[!t]
\centering
\includegraphics[width=85mm]{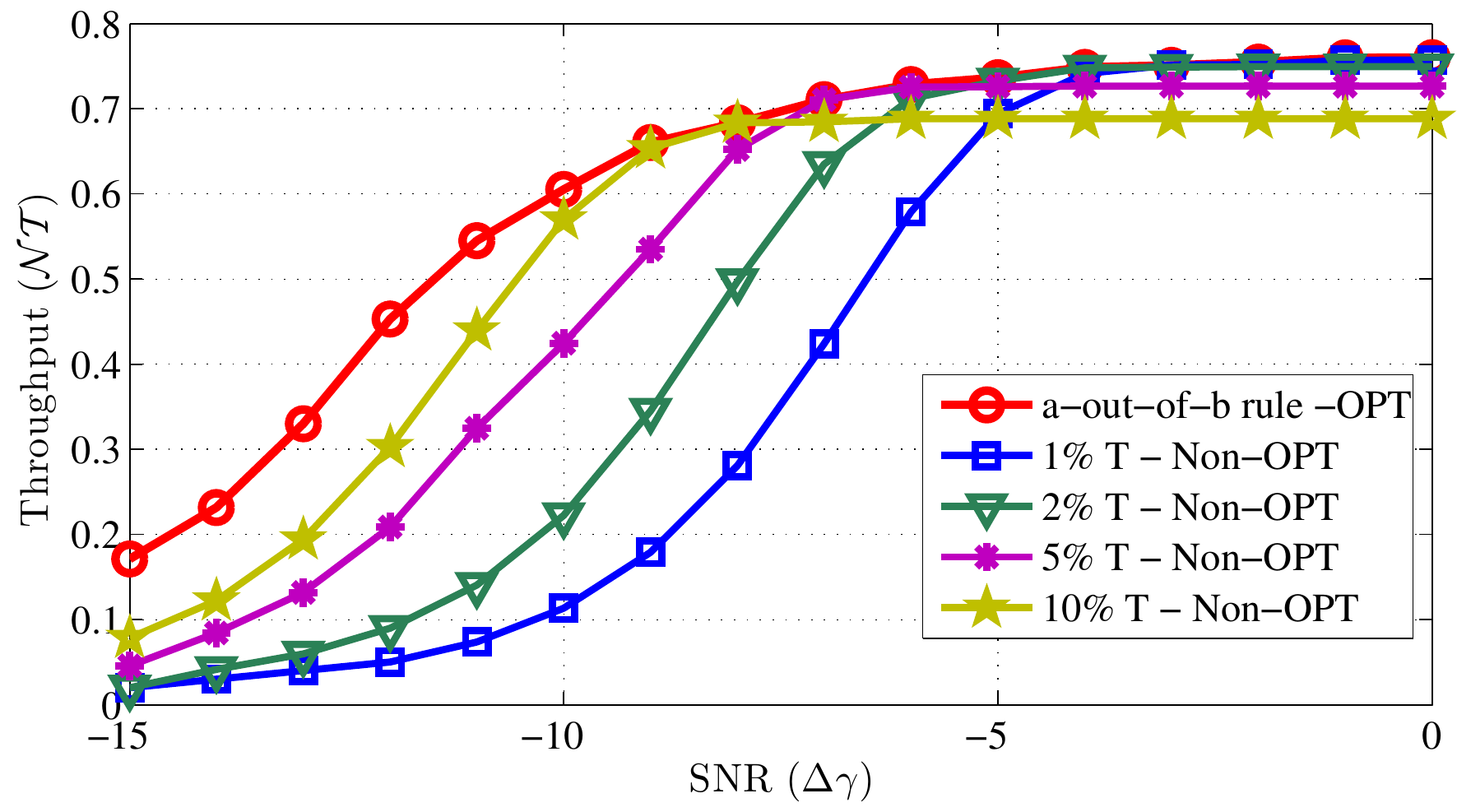}
\caption{Normalized throughput versus SNR shift $\Delta \gamma$ for $N=10$ and $M=4$ for optimized and non-optimized scenarios.}
\label{Fig6}
\end{figure}

\begin{figure}[!t]
\centering
\includegraphics[width=85mm]{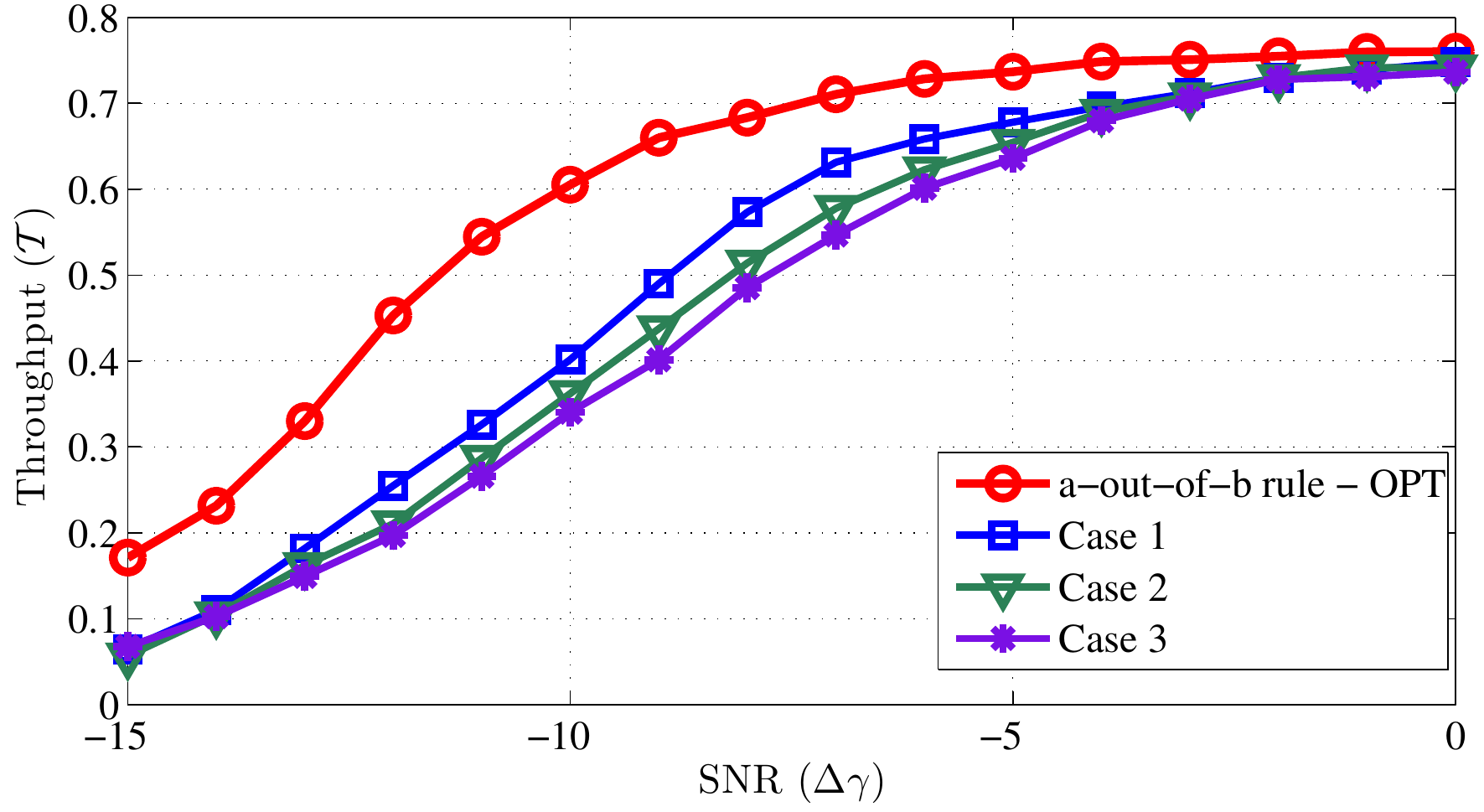}
\caption{Normalized throughput versus SNR shift $\Delta \gamma$ for $N=10$ and $M=4$ for optimized and RR channel assignments.}
\label{Fig7}
\end{figure}

To investigate the performance of our proposed algorithm for a typical system, we consider the network setting with $N=10$ and $M=4$. We divide SUs into 2 groups where
 the received SNRs at SUs due to the transmission from PU $i$ is equal
 to $\gamma^{ij}_{1,0} = -15 dB$ and $\gamma^{ij}_{2,0} = -10 dB$ (or their shifted values described later) for the two groups, respectively. 
Again, to describe the SNR of different SUs and channels, we use  $\left\{ i,j \right\}$ to denote a combination of channel $j$ and SU $i$ who senses this channel. 
The combinations of the first group corresponding to $\gamma_{1,0}^{ij} = -10 dB$ are chosen as follows: channel 1: $\left\{1,1\right\}, \left\{2,1\right\}, \left\{3,1\right\} $; channel 2: $\left\{2,2\right\}, \left\{4,2\right\}, \left\{5,2\right\}$; channel 3: $\left\{4,3\right\}, \left\{6,3\right\}, \left\{7,3\right\}$; and channel 4: $\left\{1,4\right\}, \left\{3,4\right\}, \left\{6,4\right\}, \left\{8,4\right\}, \left\{9,4\right\}, \left\{10,4\right\}$. The remaining combinations belong to the second group with
the SNR equal to $\gamma_{2,0}^{ij} = -15 dB$. To obtain results for different values of SNRs, we consider different shifted sets of SNRs  where $\gamma_{1}^{ij} $ and 
$\gamma_{2}^{ij}$ are shifted by 
$\Delta \gamma$ around their initial values $\gamma_{1,0}^{ij} = -15 dB$ and $\gamma_{2,0}^{ij} = -10 dB$ as $\gamma_{1}^{ij} = \gamma_{1,0}^{ij} + \Delta \gamma $ and $\gamma_{2}^{ij} = \gamma_{2,0}^{ij} + \Delta \gamma $. For example, as $\Delta \gamma = -10$, the resulting SNR values are $\gamma_1^{ij} = -25 dB$ and $\gamma_2^{ij} = -20 dB$. 
These parameter settings are used to obtain the results presented in Figs.~\ref{Convergence_NT_Iter},~\ref{Fig4},~\ref{Fig5},~\ref{Fig6}, and \ref{Fig7} in the following.

Fig.~\ref{Convergence_NT_Iter} illustrates the convergence of Alg. 2 where
we show the normalized throughput ${\mathcal NT}_p$ versus the iterations for $\Delta \gamma = -2, -5, -8 $ and $-11 dB$.
For simplicity, we choose $\delta $ equals $10^{-3} \times {\mathcal NT}_c $ in Alg. 2. 
This figure confirms that Alg.~\ref{ChanAAp} converges after about 11, 13, 15 and 16 iterations  for $\Delta \gamma = -2,-5, -8,$ and $-11 dB$, respectively.
In addition, the normalized throughput increases over the iterations as expected.

Fig.~\ref{Fig4} presents normalized throughput $\mathcal{NT}_p$ versus transmission probability $p$ and sensing time $\tau^{11}$ for 
the SNR shift equal to $\Delta \gamma = -7$ where the sensing times for other pairs of SUs and channels are optimized as in Alg. \ref{mainalg_pg}. 
This figure shows that channel sensing and access parameters can strongly impact the throughput of the secondary network,
which indicates the need to optimize them. This figure shows that the optimal values of  $p$ and  $\tau^{11}$ are around
 $\left({\bar \tau}^{11}, {\bar p}\right)= \left(0.0054 s, 0.1026\right)$ to achieve the maximum normalized throughput of $\mathcal{NT}_p = 0.7104 $. 
It can be observed that normalized throughput $\mathcal{NT}_p$ is less sensitive to transmission probability $p$ 
while it varies more significantly as the sensing time $\tau^{11}$ deviates from the optimal value.
In fact, there can be multiple available channels which each SU can choose from. Therefore, the contention level on each
available channel would not be very intense for most values of $p$. This explains why the throughput is not
very sensitive to the access parameter $p$.

% Fig. 5

In Fig.~\ref{Fig5}, we compare the normalized throughput of the secondary network as each SU employs four different aggregation
 rules, namely AND, OR, majority, and the optimal a-out-of-b rules. The
four throughput curves in this figure represent the optimized normalized throughput values achieved by using Algs. \ref{mainalg_pg} and \ref{ChanAAp}. 
For the OR, AND, majority rules, we do not need to find optimized $a_j$ parameters for different channels $j$ in Alg. \ref{mainalg_pg}. 
Alternatively, $a_j = 1 $, $a_j = b_j $ and $a_j = \left\lceil b/2 \right\rceil $ correspond to the OR, AND and majority rules, respectively.
It can be seen that the optimal a-out-of-b rule achieves the highest throughput among the considered rules. 
Moreover, the performance gaps between the optimal a-out-of-b rule and other rule tends to be larger for smaller SNR values.
% Fig. 6

In Fig.~\ref{Fig6}, we compare the throughput performance as the sensing times are  optimized by using Alg. \ref{mainalg_pg} and they are fixed at different fractions of the cycle time in Alg. \ref{mainalg_pg}.  
For fair comparison, the optimized a-out-of-b rules are used in both schemes with optimized and non-optimized sensing times.
For the non-optimized scheme, we employ Alg. \ref{ChanAAp} for channel assignment; however, we do not optimize the sensing times in Alg. \ref{mainalg_pg}. 
Alternatively, $\tau^{ij}$ is chosen from the following values: $1\% T$, $2\%T$, $5\%T$ and $10\%T$ where $T$ is the cycle time. 
Furthermore, for this non-optimized scheme, we still find an optimized value of ${\bar a}_j$ for each channel $j$ (corresponding to the sensing phase) and the optimal value of ${\bar p} $ (corresponding to the access phase) in Alg. \ref{mainalg_pg}. 
This figure confirms that the optimized design achieves the largest throughput.
Also, small sensing times can achieve good throughput performance at the high-SNR regime but result in poor performance if the SNR values are low.
In contrast, too large sensing times (e.g., equal $10\%T $) may become inefficient if the SNR values are sufficiently large. 
These observations again illustrate the importance of optimizing the channel sensing and access parameters.

% Fig. 7

\begin{table}[!t]
\centering
\caption{Round-robin Channel Assignment (x denotes an assignment)}
\label{table_round}
\begin{tabular}{|c|c|c|c|c|c|c|c|c|c|c|c|c|c|}
\cline{3-14} 
\multicolumn{2}{c|}{} & \multicolumn{12}{c|}{\textbf{Channel}}\tabularnewline
\cline{3-14} 
\multicolumn{1}{c}{} &  & \multicolumn{4}{c|}{\textbf{Case 1}} & \multicolumn{4}{c|}{\textbf{Case 2}} & \multicolumn{4}{c|}{\textbf{Case 3}}\tabularnewline
\cline{3-14} 
\multicolumn{2}{c|}{} & 1 & 2 & 3 & 4 & 1 & 2 & 3 & 4 & 1 & 2 & 3 & 4\tabularnewline
\hline 
 & 1 & x &  &  &  & x & x &  &  & x & x & x & \tabularnewline
\cline{2-14} 
 & 2 &  & x &  &  &  & x & x &  &  & x & x & x\tabularnewline
\cline{2-14} 
 & 3 &  &  & x &  &  &  & x & x &  &  & x & x\tabularnewline
\cline{2-14} 
 & 4 &  &  &  & x &  &  &  & x &  &  &  & x\tabularnewline
\cline{2-14} 
\textbf{SU} & 5 & x &  &  &  & x & x &  &  & x & x & x & \tabularnewline
\cline{2-14} 
 & 6 &  & x &  &  &  & x & x &  &  & x & x & x\tabularnewline
\cline{2-14} 
 & 7 &  &  & x &  &  &  & x & x &  &  & x & x\tabularnewline
\cline{2-14} 
 & 8 &  &  &  & x &  &  &  & x &  &  &  & x\tabularnewline
\cline{2-14} 
 & 9 & x &  &  &  & x & x &  &  & x & x & x & \tabularnewline
\cline{2-14} 
 & 10 &  & x &  &  &  & x & x &  &  & x & x & x\tabularnewline
\hline
\end{tabular}
\end{table}

We compare the normalized throughput under our optimized design and the round-robin (RR) channel assignment strategies in Fig.~\ref{Fig7}. 
For RR channel assignment schemes, we first allocate channels for SUs as described in Table~\ref{table_round} (i.e., we consider three different RR channel assignments). 
In the considered round-robin channel assignment schemes, we assign at most 1, 2 and 3 channels for each SU corresponding to cases 1, 2 and 3 as shown in Table~\ref{table_round}.
In particular, we sequentially assign channels with increasing indices for the next SUs until exhausting (we then repeat this procedure for the following SU).
Then, we only employ Alg. \ref{mainalg_pg} to optimize the sensing and access parameters for these RR channel assignments. 
Fig.~\ref{Fig7} shows that the optimized design achieves much higher throughput than those due to RR channel assignments. 
These results confirm that channel assignments for cognitive radios play a very important role in maximizing the spectrum utilization for CRNs.
In particular, if it would be sufficient to achieve good sensing and throughput performance if we assign a small number of nearby SUs to sense any particular channel instead of requiring all SUs to sense the channel. 
This is because ``bad SUs'' may not contribute to improve the sensing performance but result in more sensing overhead, which ultimately decreases the throughput of 
the secondary network.

% Fig. 8
\begin{figure}[!t]
\centering
\includegraphics[width=85mm]{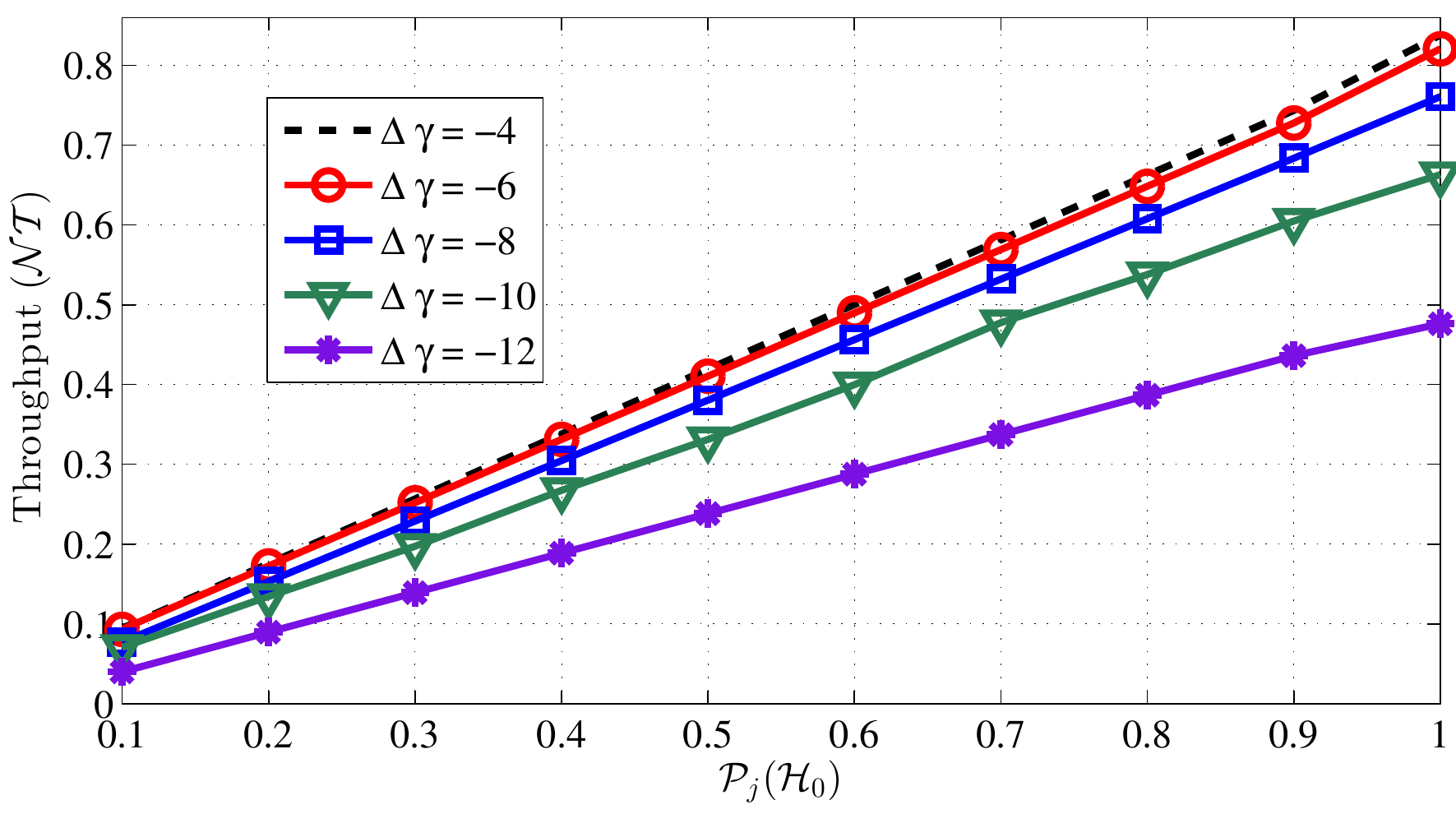}
\caption{Normalized throughput versus probability of having vacant channel $\mathcal{P}_j \left(\mathcal{H}_0\right)$ for $N=10$ and $M=4$ for optimized 
channel assignments and a-out-of-b aggregation rule.}
\label{Fig8}
\end{figure}

In Fig.~\ref{Fig8}, we consider the impact of PUs' activities on throughput performance of the secondary network. In particular, we vary the probabilities of having 
idle channels for secondary spectrum access ($\mathcal{P}_j \left(\mathcal{H}_0\right)$) in the range of $\left[0.1,1\right]$. For larger values of $\mathcal{P}_j \left(\mathcal{H}_0\right)$, there are more opportunities for SUs to find spectrum holes to transmit data, which results in higher throughput and vice versa. Moreover, this figure
shows that the normalized throughput increases almost linearly with $\mathcal{P}_j \left(\mathcal{H}_0\right)$. Also as the $\Delta \gamma$ increases (i.e., higher SNR), 
the throughput performance  can be improved significantly. However,  the improvement becomes negligible if the SNR values are sufficiently large (for $\Delta \gamma$ in $\left[-6,-4\right]$). This is because for large SNR values, the required sensing time is sufficiently small, therefore, further increase of SNR does not reduce the sensing time much to improve 
the normalized throughput.

% Fig. 9
\begin{figure}[!t]
\centering
\includegraphics[width=85mm]{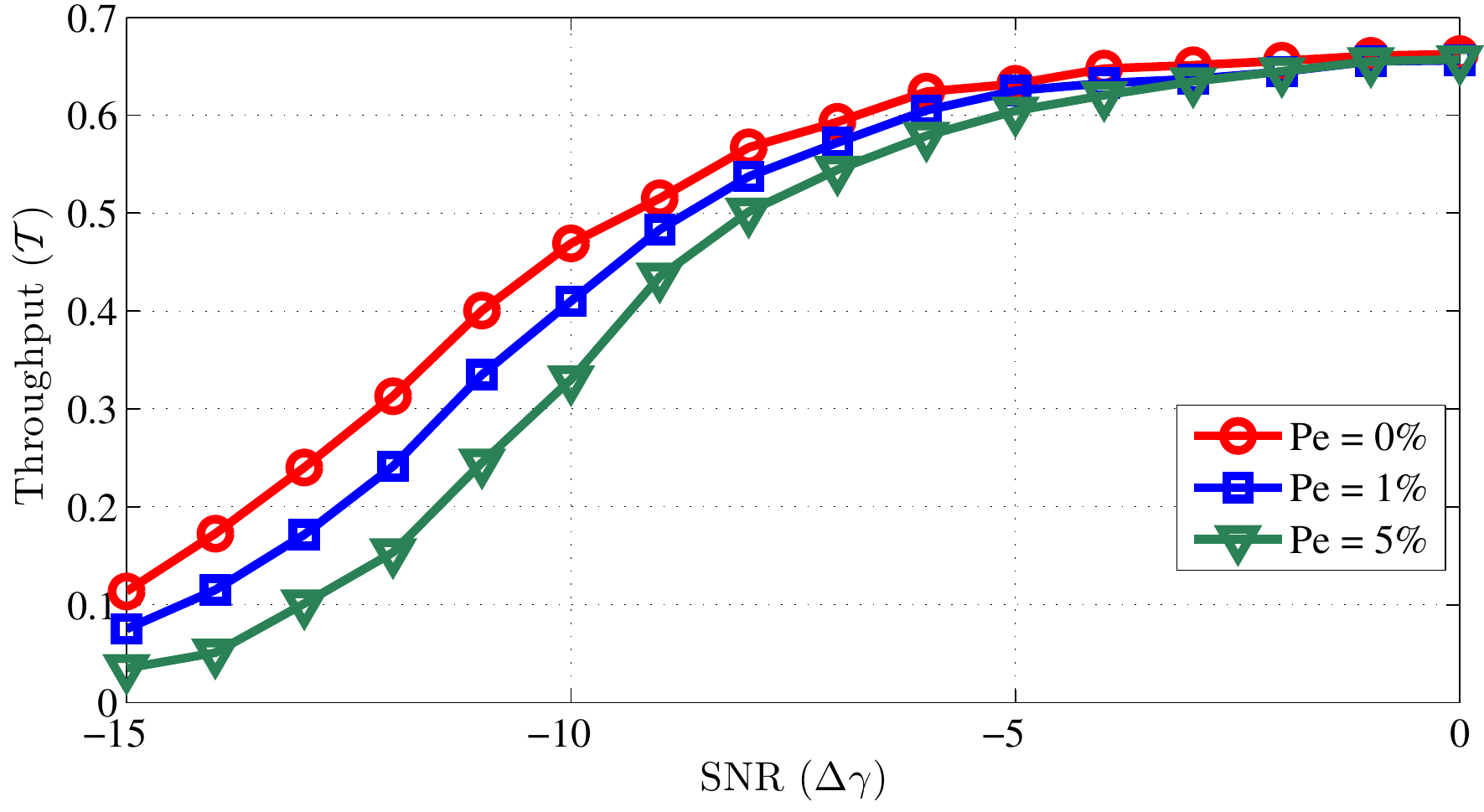}
\caption{Normalized throughput versus SNR shift $\Delta \gamma$ for $N=4$ and $M=3$ for optimized channel assignments and a-out-of-b aggregation rules.}
\label{Fig9}
\end{figure}

Finally, we study the impact of reporting errors on the throughput performance by using the extended throughput analytical model in Section ~\ref{Exten}. The network setting under
investigation has $N=4 $ SUs and $M=3 $ channels. Again, we use notation $\left\{ i,j \right\}$ to represent a combination of channel $j$ and SU $i$. 
The combinations with $\gamma_{10}^{ij} = -10 dB$ are chosen as follows: channel 1: $\left\{1,1\right\}, \left\{2,1\right\}, \left\{3,1\right\} $; channel 2: $\left\{2,2\right\}, \left\{4,2\right\}$; channel 3: $\left\{1,3\right\}, \left\{4,3\right\}$. The remaining combinations correspond to $\gamma_{20}^{ij} = -15 dB$. 
We assume that the reporting errors between every pair of 2 SUs are the same, which is denoted as $P_e$.
In Fig.~\ref{Fig9}, we show the achieved throughput as $P_e = 0\% $, $P_e = 1\% $ and $P_e = 5\% $ under optimized design. 
We can see that when $P_e$ increases, the normalized throughput decreases quite significantly if the SNR is sufficiently low.
However, in the high-SNR regime, the throughput performance is less sensitive to the reporting errors.

\vspace{0.2cm}
\section{Conclusion}
\label{conclusion} 

We have proposed a general analytical and optimization framework for SDCSS and access design in multi-channel CRNs. 
In particular, we have proposed the $p$-persistent CSMA MAC protocol integrating the SDCSS mechanism. Then,
we have analyzed the throughput performance of the proposed design and have developed an efficient algorithm to optimize its sensing and
access parameters. Moreover, we have presented both optimal brute-force search and low-complexity algorithms to determine efficient channel sensing sets 
and have analyzed their complexity. We have also extended the framework to consider reporting errors in exchanging sensing results among SUs.
Finally, we have evaluated the impacts of different parameters on the throughput performance of the proposed design
and illustrated the significant performance gap between the optimized and non-optimized designs. Specifically, it has been
confirmed that optimized sensing and access parameters as well as channel assignments can achieve considerably better throughput
performance than that due to the non-optimized design. 
In the future, we will extend SDCSS and MAC protocol design for the multihop CRNs.

\bibliographystyle{IEEEtran}

%\bibliography{am_ger_eng,rubi_eng}

\begin{table*} \label{table10}
\centering
\caption{Summary of Key Variables}
\label{Summary_var}
\scriptsize
\begin{tabular}{|c||c|c||c||c||c||c||c||c||c||c||c|}
\hline 
\multicolumn{2}{|c|}{\textbf{Variable}} & \multicolumn{10}{c|}{\textbf{Description}}\tabularnewline
\hline 
\multicolumn{12}{|c|}{\textbf{Key variables for no-reporting-error scenario}}\tabularnewline
\hline 
\multicolumn{2}{|c|}{$\mathcal{P}_j\left(\mathcal{H}_0\right)$ ($\mathcal{P}_j\left(\mathcal{H}_1\right)$)} & \multicolumn{10}{l|}{probability that channel $j$ is available (or not available)}\tabularnewline
\hline 
\multicolumn{2}{|c|}{$\mathcal{P}_d^{ij}$($\mathcal{P}_f^{ij}$)} & \multicolumn{10}{l|}{probability of detection (false alarm) experienced by SU $i$ for
channel $j$}\tabularnewline
\hline 
\multicolumn{2}{|c|}{$\mathcal{P}_d^j$ ($\mathcal{P}_f^j$)} & \multicolumn{10}{l|}{probability of detection (false alarm) for channel $j$ under SDCSS}\tabularnewline
\hline 
\multicolumn{2}{|c|}{$\varepsilon^{ij}$, $\gamma^{ij}$} & \multicolumn{10}{l|}{detection threshold, signal-to-noise ratio of the PU's signal}\tabularnewline
\hline 
\multicolumn{2}{|c|}{$\tau^{ij}$, $\tau$} & \multicolumn{10}{l|}{sensing time at SU $i$ on channel $j$, total sensing time}\tabularnewline
\hline 
\multicolumn{2}{|c|}{$N_0$, $f_s$} & \multicolumn{10}{l|}{noise power, sampling frequency}\tabularnewline
\hline 
\multicolumn{2}{|c|}{$a_j$, $b_j$} & \multicolumn{10}{l|}{parameters of $a$-out-of-$b$ rule for channel $j$}\tabularnewline
\hline 
\multicolumn{2}{|c|}{$N$, $M$} & \multicolumn{10}{l|}{total number of SUs, total number of channels}\tabularnewline
\hline 
\multicolumn{2}{|c|}{$\mathcal{S}_j^U$, $\mathcal{S}^U$} & \multicolumn{10}{l|}{set of SUs that sense channel $j$, set of all $N$ SUs}\tabularnewline
\hline 
\multicolumn{2}{|c|}{$\mathcal{S}_i$, $\mathcal{S}$} & \multicolumn{10}{l|}{set of assigned channels for SU $i$, set of all $M$ channels}\tabularnewline
\hline 
\multicolumn{2}{|c|}{$\Phi_l^k$} & \multicolumn{10}{l|}{particular set $k$ of $l$ SUs}\tabularnewline
\hline 
\multicolumn{2}{|c|}{$\Psi_{k_0}^{l_0}$} & \multicolumn{10}{l|}{set $l_0$ of $k_0$ actually available channels }\tabularnewline
\hline 
\multicolumn{2}{|c|}{$\Theta_{k_1}^{l_1}$, $\Omega_{k_2}^{l_2}$} & \multicolumn{10}{l|}{set $l_1$ of $k_1$
available channels (which are indicated by sensing outcomes), }\tabularnewline
\multicolumn{2}{|c|}{} & \multicolumn{10}{l|}{set $l_2$ of $k_2$ misdetected channels (which are indicated by sensing outcomes)}\tabularnewline
\hline 
\multicolumn{2}{|c|}{$\mathcal{NT}$} & \multicolumn{10}{l|}{normalized throughput per one channel}\tabularnewline
\hline 
\multicolumn{2}{|c|}{$\mathcal{T}_p^{ne}$, $\mathcal{T}_{j_2}^{ne}$} & \multicolumn{10}{l|}{conditional throughput: for one particular realization of sensing
outcomes corresponding to 2 sets $\Theta_{k_1}^{l_1}$ and $\Omega_{k_2}^{l_2}$,}\tabularnewline
\multicolumn{2}{|c|}{} & \multicolumn{10}{l|}{for a particular channel $j_2$}\tabularnewline
\hline 
\multicolumn{2}{|c|}{$n_j$, $k_e$} & \multicolumn{10}{l|}{number of SUs who select channel $j$ to access, $k_e=\mid\Theta_{k_1}^{l_1}\bigcup\Omega_{k_2}^{l_2}\mid$}\tabularnewline
\hline 
\multicolumn{2}{|c|}{$T$, $T_R$} & \multicolumn{10}{l|}{cycle time, total reporting time}\tabularnewline
\hline 
\multicolumn{2}{|c|}{$T_S$, $\overline{T}_S$} & \multicolumn{10}{l|}{time for transmission of packet, time for successful RTS/CTS transmission}\tabularnewline
\hline 
\multicolumn{2}{|c|}{$T_I^{i,j}$ ($\overline{T}_I^j$)} & \multicolumn{10}{l|}{$i$-th duration between 2 consecutive RTS/CTS transmission on channel $j$ (its average value)}\tabularnewline
\hline 
\multicolumn{2}{|c|}{$T_C$, $\overline{T}_{cont}^j$} & \multicolumn{10}{l|}{duration of collision, average contention time on channel $j$}\tabularnewline
\hline 
\multicolumn{2}{|c|}{$PD$} & \multicolumn{10}{l|}{propagation delay}\tabularnewline
\hline 
\multicolumn{2}{|c|}{$PS$, $ACK$} & \multicolumn{10}{l|}{lengths of packet and acknowledgment, respectively}\tabularnewline
\hline 
\multicolumn{2}{|c|}{$SIFS$, $DIFS$} & \multicolumn{10}{l|}{lengths of short time interframe space and distributed interframe space, respectively}\tabularnewline
\hline 
\multicolumn{2}{|c|}{$RTS$, $CTS$} & \multicolumn{10}{l|}{lengths of request-to-send and clear-to-send, respectively}\tabularnewline
\hline 
\multicolumn{2}{|c|}{$p$, $\mathcal{P}_C^j$} & \multicolumn{10}{l|}{transmission probability, probability of a generic slot corresponding to collision}\tabularnewline
\hline 
\multicolumn{2}{|c|}{$\mathcal{P}_S^j$, $\mathcal{P}_I^j$} & \multicolumn{10}{l|}{probabilities of a generic slot corresponding to successful transmission, idle slot}\tabularnewline
\hline 
\multicolumn{2}{|c|}{$N_c^j$($\overline{N}_c^j$)} & \multicolumn{10}{l|}{number of collisions before the first successful RTS/CTS exchange
(its average value)}\tabularnewline
\hline 
\multicolumn{2}{|c|}{$f_X^{N_c}$, $f_X^I$} & \multicolumn{10}{l|}{pmfs of $N_c^j$, $T_I^{i,j}$}\tabularnewline
\hline 
\multicolumn{12}{|c|}{\textbf{Key variables as considering reporting errors}}\tabularnewline
\hline 
\multicolumn{2}{|c|}{$\mathcal{P}_e^{i_1i_2}$} & \multicolumn{10}{l|}{probability of reporting errors between SUs $i_1$ and $i_2$}\tabularnewline
\hline 
\multicolumn{2}{|c|}{$\mathcal{P}_d^{i_1i_2j}$ ($\mathcal{P}_f^{i_1i_2j}$)} & \multicolumn{10}{l|}{probabilities of detection (false alarm) experienced by SU $i_1$ on channel $j$ with the sensing result received from SU $i_2$}\tabularnewline
\hline 
\multicolumn{2}{|c|}{$\Theta_{k_1,j_3}^{l_1}$} & \multicolumn{10}{l|}{$l_1$-th set of $k_1$ SUs whose sensing outcomes indicate that channel $j_3$ is vacant}\tabularnewline
\hline 
\multicolumn{2}{|c|}{$\Omega_{k_2,j_4}^{l_2}$} & \multicolumn{10}{l|}{$l_2$-th set of $k_2$ SUs whose sensing outcomes indicate that channel $j_4$ is vacant due to misdetection}\tabularnewline
\hline 
\multicolumn{2}{|c|}{$\Phi_{k_3,j_3}^{l_3}$} & \multicolumn{10}{l|}{$l_3$-th set of $k_3$ SUs in $\Theta_{k_1,j_3}^{l_1}$ who correctly report their sensing information on channel $j_3$ to SU $i_4$}\tabularnewline
\hline 
\multicolumn{2}{|c|}{$\Lambda_{k_4,j_3}^{l_4}$} & \multicolumn{10}{l|}{$l_4$-th set of $k_4$ SUs in $S_{j_3}^U\setminus\Theta_{k_1,j_3}^{l_1}$ who incorrectly report their sensing information on channel $j_3$ to SU $i_4$}\tabularnewline
\hline 
\multicolumn{2}{|c|}{$\Xi_{k_5,j_4}^{l_5}$} & \multicolumn{10}{l|}{$l_5$-th set of $k_5$ SUs in $\Omega_{k_2,j_4}^{l_2}$ who correctly report their sensing information on channel $j_4$ to SU $i_9$}\tabularnewline
\hline 
\multicolumn{2}{|c|}{$\Gamma_{k_6,j_4}^{l_6}$} & \multicolumn{10}{l|}{$l_6$-th set of $k_6$ SUs in $\mathcal{S}_{j_4}^U\setminus\Omega_{k_2,j_4}^{l_2}$ who incorrectly report their sensing information on channel $j_4$ to SU $i_9$}\tabularnewline
\hline 
\multicolumn{2}{|c|}{$\mathcal{S}_{1,i}^a$, $\mathcal{S}_{2,i}^a$} & \multicolumn{10}{l|}{sets of actually available channels and available due to sensing and/or
reporting errors, respectively }\tabularnewline
\hline 
\multicolumn{2}{|c|}{$\mathcal{\hat S}_1^a$, $\mathcal{\hat S}_2^a$} & \multicolumn{10}{l|}{$\mathcal{\hat S}_1^a=\bigcup_{i\in \mathcal{S}^U}\mathcal{S}_{1,i}^a$,
$\mathcal{\hat S}_2^a=\bigcup_{i\in \mathcal{S}^U}\mathcal{S}_{2,i}^a$}\tabularnewline
\hline 
\multicolumn{2}{|c|}{$\mathcal{S}_i^a$, $\mathcal{\hat S}^a$} & \multicolumn{10}{l|}{$\mathcal{S}_i^a=\mathcal{S}_{1,i}^a\bigcup \mathcal{S}_{2,i}^a$, $\mathcal{\hat S}^a=\mathcal{\hat S}_1^a\bigcup \mathcal{\hat S}_2^a$}\tabularnewline
\hline 
\multicolumn{2}{|c|}{$k_e^i$, $k_{max}$} & \multicolumn{10}{l|}{$k_e^i=\mid \mathcal{S}_i^a \mid$, $k_{max}=\mid \mathcal{\hat S}^a\mid$}\tabularnewline
\hline 
\multicolumn{2}{|c|}{$\Psi_j^a$, $\Psi^a$ } & \multicolumn{10}{l|}{set of SUs whose SDCSS outcomes indicate that channel $j$ is available,
$\Psi^a=\bigcup_{j\in \mathcal{\hat S}^a}\Psi_j^a$}\tabularnewline
\hline 
\multicolumn{2}{|c|}{$N_j$, $N_{max}$} & \multicolumn{10}{l|}{$N_j=\mid\Psi_j^a\mid$, $N_{max}=\mid\Psi^a\mid$}\tabularnewline
\hline 
\multicolumn{2}{|c|}{$\mathcal{T}_p^{re}$, $\mathcal{T}_{j_2}^{re}$} & \multicolumn{10}{l|}{conditional throughput for one particular realization of sensing
outcomes and for a particular channel $j_2$, respectively}\tabularnewline
\hline
\end{tabular}
\end{table*}

\end{document}